\documentclass[journal,10pt]{IEEEtran}

\makeatletter
\def\ps@headings{%
	\def\@oddhead{\mbox{}\scriptsize\rightmark \hfil \thepage}%
	\def\@evenhead{\scriptsize\thepage \hfil \leftmark\mbox{}}%
	\def\@oddfoot{}%
	\def\@evenfoot{}}
\makeatother 

\usepackage{CJKutf8}
\usepackage{graphicx}
\usepackage{amsmath, amsfonts,epsfig, multirow, floatflt,float}
\usepackage{amssymb}
\usepackage{cite}
\usepackage{algorithm,algorithmic}
\usepackage{hyperref}
\usepackage{enumitem}
\usepackage{mathrsfs}
\usepackage{url}
\usepackage{color}
\usepackage{subfigure}
\usepackage{array}
\usepackage{makecell}

\allowdisplaybreaks

\graphicspath{{figure/}}

\begin{document}
	
\title{Multi-Timescale Control and Communications with Deep Reinforcement Learning---Part II: Control-Aware Radio Resource Allocation}

	\author{Lei~Lei, {\it Senior Member, IEEE}, Tong~Liu, Kan~Zheng, {\it Senior Member, IEEE}, Xuemin (Sherman) Shen {\it Fellow, IEEE}}

\maketitle
\begin{abstract}
\begin{CJK}{UTF8}{gbsn}
In Part I of this {two-part} paper (Multi-Timescale Control and Communications with Deep Reinforcement Learning---Part I: Communication-Aware Vehicle Control), we decomposed the multi-timescale control and communications (MTCC) problem in Cellular Vehicle-to-Everything (C-V2X) system into a communication-aware Deep Reinforcement Learning (DRL)-based platoon control (PC) sub-problem and a control-aware DRL-based radio resource allocation (RRA) sub-problem. We focused on the PC sub-problem and proposed the MTCC-PC algorithm to learn an optimal PC policy given an RRA policy. {In this paper (Part II)}, we first focus on the RRA sub-problem in MTCC assuming a PC policy is given, and propose the MTCC-RRA algorithm to learn the RRA policy. Specifically, we incorporate the PC advantage function in the RRA reward function, which quantifies the amount of PC performance degradation caused by observation delay. Moreover, we augment the state space of RRA with PC action history for a more well-informed RRA policy. In addition, we utilize reward shaping and reward backpropagation prioritized experience replay (RBPER) techniques to efficiently tackle the multi-agent and sparse reward problems, respectively. Finally, a sample- and computational-efficient training approach is proposed to jointly learn the PC and RRA policies in an iterative process. In order to verify the effectiveness of the proposed MTCC algorithm, we performed experiments using real driving data for the leading vehicle, where the performance of MTCC is compared with those of the baseline DRL algorithms.
\end{CJK}\par
\end{abstract}

\begin{IEEEkeywords}
Multi-Timescale Decision-Making; Radio Resource Allocation; Deep Reinforcement Learning
\end{IEEEkeywords}

\section{Introduction}
In Part I of this { two-part} paper, we introduce the problem of multi-timescale control and communications (MTCC) in Cellular Vehicle-to-Everything (C-V2X) systems, where two types of decisions need to be made at different timescales: (1) platoon control (PC) decisions that are made with a coarse time grid of every $T$ $\mathrm{ms}$ control interval; (2) radio resource allocation (RRA) decisions that are made with a fine time grid of every $1$ $\mathrm{ms}$ communication interval. PC aims to determine the control inputs for following vehicles so that all vehicles move at the same speed while maintaining the desired distance between each pair of preceding vehicle (i.e., predecessor) and following vehicle (i.e., follower) \cite{li2017dynamical}. The C-V2X communications enable information exchange between vehicles so that more informed PC decisions can be made to reduce the inter-vehicle distance while guaranteeing string stability. In the C-V2X system, Vehicle-to-Vehicle (V2V) links coexist with Vehicle-to-Infrastructure (V2I) links. A V2I link connects a vehicle to the Base Station (BS) and is used for high-throughput services, and a V2V link connects a pair of predecessor and follower for periodic transmission of Collaborative Adaptive Message (CAM) according to the Predecessors Following (PF) information typology (IFT) \cite{lei2022deep}. The PC decisions are made at each follower with the target of optimizing its PC performance, which is affected by the observation delay that depends on the RRA decisions of the C-V2X system. The RRA decisions including sub-channel allocation and power control are made at each predecessor with the targets of (1) maximizing the V2I throughput; (2) minimizing the PC performance degradation due to delayed observation. The two targets are contradictory with each other and an optimal trade-off should be struck. The trade-off heavily depends on the impact of observation delay on PC performance, which in turn is affected by the PC decisions. \par

The interplay between RRA and PC necessitates the collaborative design of communications and control functions. RRA in C-V2X systems should be control-aware, taking into account the control performance degradation due to the delay or packet loss for control-related information delivery. Meanwhile, PC should be communication-aware, considering the statistical properties of random delay and packet loss in C-V2X communications.  Both PC and RRA are Sequential Stochastic Decision Problem (SSDP), where a sequence of decisions have to be made over a specific time horizon for a dynamic system whose states evolve in the face of uncertainty. We believe that tackling the SSDP under a unified Deep Reinforcement Learning (DRL) framework will better reveal the inter-dependency between control and communications, and thus facilitate the joint optimization task \cite{lei2020deep}. \par

Since employing the full-space approach to solve the multi-timescale SSDP yields formidable computation complexity, we decompose the MTCC problem into two sub-problems, i.e., (1) communication-aware DRL-based PC, and (2) control-aware DRL-based RRA. In Part I of this { two-part} paper, we have studied the communication-aware DRL-based PC assuming an RRA policy is given, and proposed the MTCC-PC algorithm to learn the PC policy. The PC problem with observation delay is essentially a Random Delay Decentralized Partially Observable Markov Decision Process (RD-Dec-POMDP). Each follower is a PC agent, which makes a local and delayed observation at each control interval and decides on its local actions to maximize its expected cumulative individual reward. We augment the PC state space with the observation delay and PC action history, and define the reward function for the augmented state to construct an augmented state Markov Decision Process (MDP). It is proved that the optimal policy for the MDP is optimal for the PC problem with delayed observation. We show by experiments that the proposed MTCC-PC algorithm outperforms the state-of-the-art communication-aware control by training in a delayed environment generated by fine-grained embedded simulation of C-V2X communications rather than by a simple stochastic delay model.\par

{This paper is Part II of the two-part paper}, in which we first focus on the control-aware DRL-based RRA problem assuming a PC policy is available, and propose the MTCC-RRA algorithm to learn the RRA policy. Different from most existing works on RRA in C-V2X systems \cite{omidshafiei2017deep,ye2019deep,yang2019intelligent,zia2019distributed,nan2021delay,liang2019spectrum,vu2020multi,xiang2021multi,zhang2022mean,parvini2023aoi,xu2023deep}, the RRA decisions are optimized taking into account the impact of observation delay on PC performance. {The main contributions of { this paper (Part II)} are summarized as follows:
\begin{itemize}
    \item \textbf{DRL-based Value of Information (VoI) Estimation and Optimization:} A systematic method is proposed for the DRL-based PC agent of each follower to estimate the VoI of any given information that are to be shared by its predecessor, where the VoI is leveraged by the DRL-based RRA agents of the predecessors to make optimal RRA decisions. Firstly, the PC performance degradation due to observation delay for each follower is quantified by the performance difference between the de-facto PC policy with observation delay and the optimal policy with undelayed observation. Then, the optimization objective of RRA is formulated by incorporating the performance degradation term. To solve such an optimization problem, it is rigorously proved that the advantage function of the optimal PC policy should be included in the RRA reward function. Moreover, the PC action history has to be included in the RRA state space to make the RRA policy more well-informed. One important advantage of this design is that the RRA decisions are made based on ``VoI per control interval", i.e., the advantage function of the optimal PC policy, which provides a finer-grained VoI compared with the existing VoI calculation methods.  

    % The RRA problem falls into the multi-agent domain similar to PC, where each predecessor is an RRA agent. However, the simple IL approach used in PC is not suitable for RRA since the agents cooperatively optimize the global expected cumulative reward, and it is hard for each agent to deduce its own contribution to the global reward.
    \item \textbf{Efficient DRL solution addressing multi-agent and sparse reward problems: } Since the RRA agents cooperatively optimize the global expected cumulative reward, and it is hard for each agent to deduce its own contribution to the global reward, the simple independent learner (IL) approach used in PC is not suitable for the RRA problem falling into the multi-agent domain. To solve this problem, we apply the reward shaping technique to design an individual RRA reward for each RRA agent. In addition, since the RRA rewards that reflect the PC performance degradation are only provided at the last communication interval of each control interval, it is difficult to learn an efficient RRA policy due to the sparse reward problem coupled with long-horizon planning. To solve this problem, we use the reward backpropagation prioritized experience replay (RBPER) technique\cite{zhong2017reward} to improve training efficiency.
    % Finally, the proposed MTCC-PC (resp. MTCC-RRA) algorithm depends on a given RRA (resp. PC) policy as input.
    \item \textbf{Sample- and computational-efficient training approach:} Since it is impossible to learn the optimal PC policy or RRA policy first without knowing the other optimal policy due to the interactions between PC and RRA, we design a sample- and computational-efficient training approach to jointly train MTCC-PC and MTCC-RRA in an iterative process to learn the optimal PC and RRA policies. 
\end{itemize}\par }

The rest of this paper is organized as follows. We recall the system model for the MTCC problem and the MTCC-PC algorithm in Section II. Subsequently, Sections III and IV introduce the control-aware DRL model for RRA and the MTCC-RRA algorithm. Then, a joint optimization algorithm for MTCC is presented in Section V. Section VI conducts the performance experiments, demonstrating that our proposed algorithm is superior to the baseline algorithms. Finally, Section VII concludes the paper.

\section{System Model and previous results} 
\subsection{Multi-timescale Decision-Making Framework}
Consider a platoon with a number of $N>2$ vehicles, i.e., $\mathcal{V}=\{0,1,\cdots, N-1\}$. All vehicles communicate with one another using C-V2X communications. The PC problem is considered within a finite time horizon, which is discretized into $K$ equal-length control intervals indexed by $k \in \mathcal{K} = \{0,1,\cdots, K-1\}$. The duration of each control interval is $T$ milliseconds ($\rm ms$). At each control interval $k$, the PC module of each follower $i\in\mathcal{V}\backslash \{0\}$ determines the vehicle control input $a^{\rm CL}_{i,k}$ based on the observations of the system state. The vehicle driving status is sampled at time $kT$. We adopt the PF IFT, where the CAM $c_{i-1,k}=\{p_{i-1,k},v_{i-1,k},acc_{i-1,k}\}$ of the predecessor $i-1\in\mathcal{V}\backslash \{N-1\}$ are transmitted to the follower $i$. For this purpose, each control interval $k$ is further divided into $T$ communication intervals of every $1$ $\rm ms$ indexed by $t \in \mathcal{T}=\{0,1,..., T-1\}$ on a faster timescale. Dynamic scheduling is considered, where the C-V2X communication module makes RRA decisions to transmit CAM at each communication interval $(k,t)$.

Since the PC decisions are made with a coarse time grid of every $T$ $\rm ms$, while the RRA decisions are made with a fine time grid of every $1$ $\rm ms$, we have a multi-timescale decision-making problem.

\subsection{Platoon Control Module and C-V2X Communications Module}
For the platoon control module, each vehicle $i\in\mathcal{V}$ obeys the dynamics model described by a first-order system, and the dynamics model is derived in discrete time on the basis of forward Euler discretization. 
	\begin{equation}
	\label{eq2}
	{p}_{i,k+1}={p}_{i,k}+Tv_{i,k},
	\end{equation}
	\begin{equation}
	\label{eq3}
	{v}_{i,k+1}={v}_{i,k}+Tacc_{i,k},
	\end{equation}
	\begin{equation}
	\label{eq4}
	{acc}_{i,k+1}=(1-\frac{T}{\tau_{i}})acc_{i,k}+\frac{T}{\tau_{i}}a^{\rm CL}_{i,k},
	\end{equation}
\noindent where $\tau_{i}$ is a time constant representing driveline dynamics. 
The tracking errors, i.e., gap-keeping error $e_{pi,k}$ and velocity error $e_{vi,k}$ of follower $i$ are defined as
	\begin{equation}
	\label{eq7}
	e_{pi,k}=d_{i,k}-d_{r,i,k},
	\end{equation}
	\begin{equation}
	\label{eq8}
	e_{vi,k}=v_{i-1,k}-v_{i,k},
	\end{equation}	
\noindent where $d_{i,k}=p_{i-1,k}-p_{i,k}-L_{i-1}$ is the headway of follower $i$ at control interval $k$, and $d_{r,i,k}=r_{i}+h_{i}v_{i,k}$ is the desired headway.

For C-V2X communications module, we consider a typical urban C-V2X network, {where $M$ V2I links (uplink considered) coexist with $N-1$ V2V links.} A V2I link $m\in \mathcal{M}=\{0,\cdots, M-1\}$ connects a vehicle to the Base Station (BS) and is used for high-throughput services. According to the PF IFT, a V2V link $i \in \mathcal{V}\backslash\{N-1\}$ connects a pair of predecessor $i$ and follower $i+1$ for periodic transmission of CAM. In order to enhance spectrum utilization, the V2V links reuse the sub-channels of the V2I links for CAM transmission. We use the binary allocation indicator $\theta_{i,m,(k,t)}\in\{0,1\}$ to indicate whether V2V link $i$ occupies sub-channel $m$ at communication interval $(k, t)$ or not. In each communication interval $(k,t)$, each vehicle $i$ transmits the data in its queue according to the local RRA decisions. 
Assume the instantaneous channel gain of V2V link $i$ over sub-channel $m$ (occupied by V2I link $m$) at communication interval $(k,t)$ is denoted by $G_{i,m,(k,t)}$. Similarly, let $G_{m,(k,t)}$ denote the channel gain of the V2I link $m$; $G_{i,B,m,(k,t)}$ the interference channel gain from V2V link $i$ transmitter to V2I link $m$ receiver; $G_{B,i,m,(k,t)}$ the interference channel gain from V2I link $m$ transmitter to V2V link $i$ receiver; and $G_{j,i,m,(k,t)}$ the interference channel gain from the V2V link $j$ transmitter to the V2V link $i$ receiver over the sub-channel $m$. The transmit power of V2V link $i$ is $P^{\rm V}_{i,m,(k,t)}$ and the constant transmit power of V2I link $m$ is $P^{\rm I}_{m}$. The SINR $\gamma_{m,(k,t)}$ of V2I link $m$ and the SINR $\gamma_{i,m,(k,t)}$ of V2V link $i$ on sub-channel $m$ at communication interval $(k,t)$ are derived by 
\begin{align}\label{SINRV2I}
\gamma_{m,(k,t)} = \frac{P^{\rm I}_{m} G_{m,(k,t)}}{\sigma^2 + \sum\limits_{i\in\mathcal{V}\backslash\{N-1\}}\theta_{i,m,(k,t)} P^{\rm V}_{i,m,(k,t)} G_{i,B,m,(k,t)}},
\end{align}
and
\begin{align}\label{SINRV2V}
\gamma_{i,m,(k,t)}= \frac{P^{\rm V}_{i,m,(k,t)} G_{i,m,(k,t)}}{\sigma^2 + I_{i,m,(k,t)}},
\end{align}
respectively,
\noindent where $\sigma^2$ is the power of channel noise which satisfies the independent Gaussian distribution with a zero mean value. $I_{i,m,(k,t)}$ is the total interference power received by V2V link $i$ over sub-channel $m$, where
\begin{align}\label{IV2V}
& I_{i,m,(k,t)}=\IEEEnonumber \\
& P^{\rm I}_{m} G_{B,i,m,(k,t)}+ \sum\limits_{j\in \mathcal{V} \backslash \{i,N-1\}}\theta_{j,m,(k,t)}  P^{\rm V}_{j,m,(k,t)} G_{j,i,m,(k,t)}.\IEEEnonumber
\end{align}

Then, the instantaneous data rate in terms of CAM of V2V link $i$ and the instantaneous data rate of V2I link $m$ at communication interval $(k,t)$ are respectively derived as
\begin{equation}\label{rateV2V}
    r^{\rm CAM}_{i,(k,t)}=\frac{\sum_{m=0}^{M-1} W\log_2(1+\gamma_{i,m,(k,t)})}{N_c},
\end{equation}
and
\begin{align}\label{rateV2I}
r_{m,(k,t)} = W\log_2(1+\gamma_{m,(k,t)} ),
\end{align}
\noindent where $W$ is the bandwidth of a sub-channel, and $N_c$ is the constant CAM size.

Let $q^{\rm CAM}_{i,(k,t)}$ denote the CAM queue length of vehicle $i$ at communication interval $(k,t)$. If the queue length $q^{\rm CAM}_{i,(k,t)}$ {reaches} the buffer capacity $N_Q$, the subsequent arriving data will be dropped. The queue process evolves as 
% \begin{figure*}
\begin{align}
    \label{queue}
    \setlength{\arraycolsep}{1.6pt}
        q^{\rm CAM}_{i,(k,t+1)}=
         \left\{
        \begin{array}{ll}
	\min \left[N_Q, \max [0,q^{\rm CAM}_{i,(k,t)}- 10^{-3}\times \right. \\ \left.r^{\rm CAM}_{i,(k,t)}]+1 \right], &\mathrm{if}\quad t=0 \\
	\max \left[0,q^{\rm CAM}_{i,(k,t)}- 10^{-3}\times r^{\rm CAM}_{i,(k,t)}\right], & \mathrm{otherwise} \\
	\end{array}\right. . 
\end{align}
% \end{figure*}
The CAM that is not transmitted during control interval $k$ will continue to be transmitted in the next control interval $k+1$.
\subsection{Correlation between Platoon Control Decisions and Radio Resource Allocation Decisions}

% In our system model, each vehicle $i\in\mathcal{V}\backslash\{0\}$ makes PC decisions on the control input $a^{\rm CL}_{i,k}$ at every control interval $k\in\mathcal{K}$. Moreover, each vehicle $i\in\mathcal{V}\backslash\{N-1\}$ makes RRA decisions on sub-channel allocation $\{\theta_{i,m,(k,t)}\}_{m\in\mathcal{M}}$ and transmit power $\{P^{\rm V}_{i,m,(k,t)}\}_{m\in\mathcal{M}}$ at every communication interval $(k,t)$, where $k\in\mathcal{K}$ and $t\in\mathcal{T}$. 
Let $\tau_{i,k}$ be the observation delay of follower $i$ at control interval $k$. Thus, $c_{i-1,k-\tau_{i,k}}=\{p_{i-1,k-\tau_{i,k}},v_{i-1,k-\tau_{i,k}},acc_{i-1,k-\tau_{i,k}}\}$ is the most recent available delayed CAM at follower $i$, which correspond to the position, velocity, and acceleration sampled at predecessor $i-1$ in control interval $k-\tau_{i,k}$. Therefore, the observed driving status of vehicle $i$ is defined as
\begin{align}
x_{i,k-\tau_{i,k}}=&\{e_{pi,k-\tau_{i,k}},e_{vi,k-\tau_{i,k}},acc_{i,k-\tau_{i,k}},acc_{i-1,k-\tau_{i,k}} \},
\end{align}  
\noindent where $e_{pi,k-\tau_{i,k}}=p_{i-1,k-\tau_{i,k}}-p_{i,k-\tau_{i,k}}-L_{i-1}-d_{r, i,k}$ and $e_{vi,k-\tau_{i,k}}=v_{i-1,k-\tau_{i,k}}-v_{i,k-\tau_{i,k}}$.
% \begin{equation}
% e_{pi,k-\tau_{i,k}}=p_{i-1,k-\tau_{i,k}}-p_{i,k-\tau_{i,k}}-L_{i-1}-d_{r, i,k}, \nonumber
% \end{equation}
% \begin{equation}
% e_{vi,k-\tau_{i,k}}=v_{i-1,k-\tau_{i,k}}-v_{i,k-\tau_{i,k}}. 
% \end{equation}

The observation delay $\tau_{i,k}$ depends on the transmission delay of CAM over V2V link $i-1$, which can be derived from $q^{\rm CAM}_{i-1,(k-1,T)}$ or $q^{\rm CAM}_{i-1,(k,0)}$ as
\begin{equation}
\label{eq23}
\tau_{i,k}=\lceil q^{\rm CAM}_{i-1,(k-1,T)} \rceil+1=\lceil q^{\rm CAM}_{i-1,(k,0)} \rceil+1.
\end{equation}

A detailed analysis of the correlation between PC and RRA decisions can be found in Part I of this { two-part} paper. 

%We jointly optimize the multi-timescale PC and RRA decisions using DRL, fully considering the interplay between them. The communication-aware PC is introduced in Part I of this paper and will be summarized in Section II.D. The control-aware RRA and the joint learning approach will be presented in Section III-V.

\subsection{Communication-Aware PC and the Corresponding DRL Solution}

We assume that the RRA policy $\pi^{\rm CM}$ is available and focus on learning the PC policy $\pi^{\rm CL}_{i}, i\in\mathcal{V}\backslash\{0\}$.

\subsubsection{{Communication-aware} PC model} 
\paragraph{PC state}	

The state for each PC agent $i$ at control interval $k$ is defined as 
\begin{equation}
\label{pc_state}
  S^{\rm CL}_{i,k}=\{x_{i,k-\tau_{i,k}},\{a^{\rm CL}_{i,k'}\}_{k'=k-\tau_{\mathrm{max}}}^{k-1},\tau_{i,k}\}.  
\end{equation}  \par

\paragraph{PC action}	
The control input, $a^{\rm CL}_{i,k}$ of the PC agent $i$ is regarded as its PC action at control interval $k$. 
\paragraph{PC reward function}	
The individual reward for each PC agent $i$ is given by 	
\begin{align}
\label{PC_Reward}
&	R^{\rm CL}_{i,k}(x_{i,k},a^{\rm CL}_{i,k})= \IEEEnonumber \\
&-\{|\frac{e_{pi,k}}{\hat{e}_{p,\mathrm{max}}}|+{\alpha_1}|\frac{e_{vi,k}}{\hat{e}_{v,\mathrm{max}}}|+{\alpha_2}|\frac{a^{\rm CL}_{i,k}}{a^{\rm CL}_{\mathrm{max}}}|+{\alpha_3}|\frac{j_{i,k}}{2acc_{\mathrm{max}}/T}|\},
\end{align}
 where $j_{i,k}=\frac {acc_{i,k+1}-acc_{i,k}}{T}=-\frac{1}{\tau_{i}}acc_{i,k}+\frac{1}{\tau_{i}}a^{\rm CL}_{i,k}$ is the jerk. \par

The objective of the PC problem is for each PC agent $i$ to find the optimal policy $\pi^{\rm CL*}_{i}$ under delayed observation that maximizes its individual expected return $J^{\rm CL}_{i}$, i.e.,
\begin{equation}
		\label{eq16}
		\pi^{\rm CL*}_{i}=\arg\max_{\pi^{\rm CL*}_{i}}J^{\rm CL}_{i}, \ \forall i\in\mathcal{V}\backslash\{0\}.
	\end{equation}   
\noindent where 
\begin{equation}
\label{pc_perf}
J^{\rm CL}_{i}=\mathrm{E}_{\pi^{\rm CM}}\mathrm{E}_{\pi^{\rm CL}_{i}}[\sum_{k=0}^{K-1} \gamma^{k} R^{\rm CL}_{i,k}], \ 0 \leq \gamma \leq 1. 
\end{equation}
Note that $\gamma$ is the PC reward discount factor. \par
%Since $\pi^{\rm CM}$ has an important influence on the state transition probabilities of the communication-aware PC model, the PC model should be trained by a delayed environment generated by the fine-grained RRA environment. 

\subsubsection{DRL {solution} for communication-aware PC model}	

%The DDPG algorithm \cite{lillicrap2015continuous} is utilized to solve the PC problem. We discuss the adoption of DDPG in the multi-agent setting and random delay setting to prove that the DDPG algorithm is suitable for solving the RD-Dec-POMDP problem of the proposed communication-aware PC model. \par

\paragraph{Multi-agent problem in DRL-based PC}
The PC problem corresponds to a Dec-POMDP and lies in the multi-agent domain. We adopt the IL approach where each agent learns independently since the non-stationary environment issue for IL is trivial in the PC problem and the credit assignment issue in the multi-agent problem does not exist for our PC model.

\paragraph{Random observation delay problem in DRL-based PC}
We approximately consider that the undelayed driving status $x_{i,k}$ at PC agent $i$ is Markov, ignoring the impact of the predecessors' actions on $x_{i,k+1}$. However, each PC agent $i$ can only observe the delayed driving status $x_{i,k-\tau_{i,k}}$ instead of $x_{i,k}$. It is proved in Theorem 1 of Part I that $S_{i,k}^{\mathrm{CL}}$ becomes a Markov state by augmenting the delayed observation of driving status with action history.\par

%The reward function $R^{\rm CL}_{i,k}(x_{i,k},a^{\rm CL}_{i,k})$ defined in \eqref{PC_Reward} is a function of the undelayed observation $x_{i,k}$ instead of $S_{i,k}^{\rm CL}$. 

We construct an MDP $\tilde{\mathcal{M}}_{i}=(S_{i,k}^{\rm CL},a^{\rm CL}_{i,k},\tilde{R}^{\rm CL}_{i,k},p,\gamma)$ for the delayed observations, where the reward function $\tilde{R}^{\rm CL}_{i,k}(S_{i,k}^{\rm CL},a^{\rm CL}_{i,k})$ for the augmented state of each follower $i\in\mathcal{V}\backslash\{0\}$ is defined as $\tilde{R}^{\rm CL}_{i,k}(S_{i,k}^{\rm CL},a^{\rm CL}_{i,k})=\mathrm{E}_{x_{i,k}}[R^{\rm CL}_{i,k}(x_{i,k},a^{\rm CL}_{i,k})|S_{i,k}^{\rm CL}]$. It is proved in Theorem 2 of Part I that the optimal policy $\tilde{\pi}^{\rm CL*}_{i}$ for the augmented state MDP $\tilde{\mathcal{M}}_{i}$ is the same as the optimal policy $\pi^{\rm CL*}_{i}$ in \eqref{eq16} for our PC problem under delayed observation.\par

%the corresponding state-value function
%\begin{equation}
%\tilde{V}^{\rm CL}_{i,k}(S_{i,k}^{\rm CL})=E[V^{\rm CL}_{i,k}(x_{i,k})|S_{i,k}^{\rm CL}].
%\end{equation}

\paragraph{MTCC-PC Algorithm}
Based on the above discussion, the DDPG algorithm \cite{lillicrap2015continuous} is utilized to solve the PC problem. The Deterministic Policy Gradient (DPG) Theorem can be directly applied since $\tilde{\mathcal{M}}_{i}$ is an MDP, i.e., the deterministic policy gradient for the augmented state MDP $\tilde{\mathcal{M}}_{i}$ is
\begin{align}
\label{DPG}
\bigtriangledown_{\theta^{\mu}_{i}}J_{i}^{\rm CL}(\mu^{\rm CL}_{{i}})&=\mathrm{E}[\bigtriangledown_{\theta^{\mu}_{i}}\mu^{\rm CL}_{i}(S^{\rm CL}_{i,k}|\theta^{\mu}_{i}) \IEEEnonumber \\
&\bigtriangledown_{a}Q^{\rm CL}_{i}(S^{\rm CL}_{i,k}, a |\theta^{Q}_{i})|_{a=\mu^{\rm CL}_{i}(S^{\rm CL}_{i,k}|\theta^{\mu}_{i}) }].
\end{align}

In order to sample the deterministic policy gradient in \eqref{DPG}, we need to evaluate the action-value function $Q_{\mu_{\theta_{i}}}^{\rm CL}(S^{\rm CL}_{i,k},a^{\rm CL}_{i,k})$ of the augmented state MDP $\tilde{\mathcal{M}}_{i}$. Based on the Bellman equation, the expected cumulative discounted reward from control interval $k$ is defined as \begin{align}
\label{augmentQ}
&Q_{\mu_{\theta_{i}}}^{\rm CL}(S^{\rm CL}_{i,k},a^{\rm CL}_{i,k})=\mathrm{E}_{x_{i,k}}[R^{\rm CL}_{i,k}(x_{i,k},a^{\rm CL}_{i,k})|S_{i,k}^{\rm CL}]  \IEEEnonumber \\
&+\gamma \mathrm{E}_{S_{i,k+1}^{\mathrm{CL}}}[Q_{\mu_{\theta_{i}}}^{\rm CL}(S^{\rm CL}_{i,k+1},\mu_{\theta_{i}}(S^{\rm CL}_{i,k+1}))|S_{i,k}^{\mathrm{CL}},a^{\rm CL}_{i,k}].
\end{align}

During training, the sampled deterministic policy gradient ascent on $Q^{\rm CL}_{i}(S_{i,k},\mu^{\rm CL}_{i}(S_{i,k}|\theta^{\mu}_{i})|\theta^{Q}_{i})$ with regard to $\theta^\mu_{i}$ is used to train the actor network, and the critic network is trained by minimizing the Root Mean Square Error (RMSE) $L_{i,k} = y_{i,k}-Q^{\rm CL}_{i}(S^{\rm CL}_{i,k}, a^{\rm CL}_{i,k} |\theta^{Q}_{i})$ using the sampled gradient descent with respect to $\theta^{Q}_{i}$, where the TD target $y_{i,k}$ is calculated by 
\begin{equation}
\label{TD}
y_{i,k}=R^{\rm CL}_{i,k}(x_{i,k},a^{\rm CL}_{i,k})+\gamma Q_{\mu_{\theta_{i}}}^{\rm CL}(S^{\rm CL}_{i,k+1},\mu_{\theta_{i}}(S^{\rm CL}_{i,k+1})).
\end{equation} \par

{Note that in the last control interval $K-1$ of the control episode, we still calculate $y_{i,k}$ using \eqref{TD} instead of only the immediate reward. In this way, we convert the finite horizon SSDP to an infinite horizon one, since DDPG and most DRL algorithms are designed for infinite or indefinite horizon problems \cite{9951132}.}\par

It is assumed that the PC agent $i$ can sample the undelayed reward $R^{\rm CL}_{i,k}(x_{i,k},a^{\rm CL}_{i,k})$ during training. This is possible since learning can be performed in a simulator or a laboratory in which the undelayed reward is available.\par

An important characteristic of the proposed MTCC-PC algorithm is that it is trained in a delayed environment generated by the simulation of C-V2X communications with de facto RRA policy rather than by a coarse-grained stochastic delay model. This is to ensure the delay distribution in the training environment is the same as that in the execution environment. 

\section{Control-Aware DRL-based Radio Resource Allocation}
We assume that the PC policy is available and focus on learning the RRA policy in this section.

The RRA problem is essentially a Dec-POMDP. Each predecessor $i\in\mathcal{V}\backslash\{N-1\}$ is an RRA agent, which makes a local observation at each communication interval $(k,t)$, and decides on its local actions to maximize the expected cumulative \emph{global} reward. Since one of the objectives of RRA is to minimize the PC performance degradation due to delayed observation, the main challenges in formulating the DRL model are the design of the reward function and state space.  \par 

\subsection{PC Performance Degradation Due to Delayed Observation}
We quantify the PC performance degradation for each follower $i\in\mathcal{V}\backslash\{0\}$ by the performance difference between $\pi_{i}^{\mathrm{CL}}(S^{\rm CL}_{i,k})$ and $\hat{\pi}_{i}^{\mathrm{CL}*}(x_{i,k})$, where the former is the policy in Section II when the PC agent only has access to the augmented state $S^{\rm CL}_{i,k}$ with delayed observation; while the latter is the optimal policy when the PC agent has oracle access to the current driving status $x_{i,k}$ at each control interval $k$. Therefore, the total PC performance degradation $\Delta J^{\mathrm{CL}}$ over all the followers is 
\begin{equation}
\label{perf_loss}
\Delta J^{\mathrm{CL}}=\sum_{i=0}^{N-2}\left(J^{\rm CL}_{i+1}-\hat{J}^{\rm CL*}_{i+1}\right),
\end{equation}
\noindent where $J^{\rm CL}_{i+1}$ is the expected return of policy $\pi_{i+1}^{\mathrm{CL}}(S^{\rm CL}_{i+1,k})$ given in \eqref{pc_perf}, and
\begin{equation}
\label{perf_truestate}
\hat{J}^{\rm CL*}_{i+1}=\mathrm{E}_{\hat{\pi}^{\rm CL*}_{i+1}}\left[\sum_{k=0}^{K-1}\gamma^{k} R^{\rm CL}_{i+1,k}\right]
\end{equation}
\noindent is the expected return of optimal policy $\hat{\pi}_{i}^{\mathrm{CL}*}(x_{i+1,k})$, which is not affected by $\pi^{\rm CM}$ since it is assumed that the PC agent can observe the current driving status $x_{i+1,k}$ without delay.

\subsection{RRA Reward Modeling}
The optimization objective of RRA is formulated as 
\begin{equation}
\label{opti_obj}
J^{\mathrm{CM}}=\kappa_1 \sum_{m=0}^{M-1}\mathrm{E}_{\pi^{\rm CM}}\left[\sum_{k=0}^{K-1}\sum_{t=0}^{T-1} \eta^{kT+t}r_{m,(k,t)}\right]+\kappa_2 \Delta J^{\mathrm{CL}},
\end{equation}
\noindent where $\eta$ is the RRA reward discount factor. The first term corresponds to the discounted sum throughput of all V2I links $m\in\mathcal{M}$ over all the communication intervals of a control episode, and the second term corresponds to the PC performance degradation given in \eqref{perf_loss}. The weight factors $\kappa_1$ and $\kappa_2$ indicate the relative importance of minimizing PC performance degradation versus maximizing the V2I throughput.\par

The optimization objective $J^{\mathrm{CM}}$ should correspond to the expected return of the DRL model, i.e.,
\begin{equation}
\label{RRA_cumreward}
J^{\mathrm{CM}}=\mathrm{E}_{\pi^{\rm CM}}\mathrm{E}_{\pi^{\rm CL}}\left[\sum_{k=0}^{K-1}\sum_{t=0}^{T-1} \eta^{kT+t} R^{\rm CM}_{(k,t)}\right], 0 \leq \eta \leq 1,
\end{equation}
\noindent where the expectation is taken with respect to the probability distribution of the state-action trajectories when the RRA policy is $\pi^{\rm CM}$ and PC policy is $\pi^{\mathrm{CL}}$. Since both RRA and PC are multi-agent problems, the RRA (resp. PC) policy corresponds to the set of policies of all the RRA (resp. PC) agents, i.e., $\pi^{\rm CM}=\{\pi_{i}^{\rm CM}\}_{i\in\mathcal{V}\backslash\{N-1\}}$ and $\pi^{\rm CL}=\{\pi_{i}^{\rm CL}\}_{i\in\mathcal{V}\backslash\{0\}}$. \par

We will try to derive the RRA reward function $R^{\rm CM}_{(k,t)}$ from \eqref{opti_obj} and \eqref{RRA_cumreward}. Specifically, we have  

\begin{align}
\label{comreward_g}
R^{\rm CM}_{(k,t)}=\left\{
\begin{array}{ll}
 \kappa_1 R_{{\rm I},(k,t)}, &  \ 0 \le t< T-1\\
 \kappa_1 R_{{\rm I},(k,t)}+\kappa_2 R_{{\rm V},(k,T)} , & \  t= T-1 \\
\end{array}\right. ,
\end{align}
\noindent where
\begin{equation}
\label{comreward_i}
R_{{\rm I},(k,t)} =\sum_{m=0}^{M-1} r_{m,(k,t)}
\end{equation}
\noindent is the component related to V2I throughput, and $R_{{\rm V},(k,T)}$ is the component related to the PC performance degradation. \par

At first glance, $R_{{\rm V},(k,T)}$ seems to be the difference between the two PC rewards $R^{\rm CL}_{i+1,k+1}(x_{i+1,k+1},\pi_{i+1}^{\mathrm{CL}}(S^{\rm CL}_{i+1,k+1}))$ and $R^{\rm CL}_{i+1,k+1}(x_{i+1,k+1},\hat{\pi}_{i+1}^{\mathrm{CL}*}(x_{i+1,k+1}))$ based on \eqref{perf_loss}, \eqref{pc_perf}, and \eqref{perf_truestate}. However, this is not correct since the expectations in \eqref{pc_perf} and \eqref{perf_truestate} are with respect to different trajectory distributions. In order to derive $R_{{\rm V},(k,T)}$, the following lemma is used.

\newtheorem{lemma}{Lemma}
\begin{lemma}
The PC performance degradation due to observation delay for follower $i$ equals to the expected cumulative advantage of $\pi_{i}^{\mathrm{CL}}(S^{\rm CL}_{i,k})$ over $\hat{\pi}_{i}^{\mathrm{CL}*}(x_{i,k})$, i.e.,
\begin{align}
\label{perf_diff}
J^{\rm CL}_{i}-\hat{J}^{\rm CL*}_{i}=\mathrm{E}_{\pi^{\rm CM}}\mathrm{E}_{\pi^{\rm CL}_{i}}\left[\sum_{k=0}^{K-1} \gamma^{k} A_{\hat{\pi}^{\rm CL*}_{i}}\left(x_{i,k},\pi^{\rm CL}_{i}(S^{\rm CL}_{i,k})\right)\right],&\IEEEnonumber \\
\ 0 \leq \gamma \leq 1,&
\end{align}
\noindent where 
\begin{align}
\label{advantage}
&A_{\hat{\pi}^{\rm CL*}_{i}}\left(x_{i,k},\pi^{\rm CL}_{i}(S^{\rm CL}_{i,k})\right)\IEEEnonumber \\ &=Q_{\hat{\pi}^{\rm CL*}_{i}}\left(x_{i,k},\pi^{\rm CL}_{i}(S^{\rm CL}_{i,k})\right)
-V_{\hat{\pi}^{\rm CL*}_{i}}(x_{i,k})
\end{align}
\noindent is the advantage function of policy $\hat{\pi}_{i}^{\mathrm{CL}*}(x_{i,k})$.
\end{lemma}

The proof of Lemma 1 is given in Appendix A.

Based on Lemma 1, $R_{{\rm V},(k,T)}$ is defined as
\begin{align}
\label{comreward_v}
R_{{\rm V},(k,T)} = &\sum_{i=0}^{N-2} A_{\hat{\pi}^{\rm CL*}_{i+1}}\left(x_{i+1,k+1},\pi^{\rm CL}_{i+1}(S^{\rm CL}_{i+1,k+1})\right)
\end{align}

{Since the advantage function $A_{\hat{\pi}^{\rm CL*}_{i+1}}\left(x_{i+1,k},\pi^{\rm CL}_{i+1}(S^{\rm CL}_{i+1,k})\right)$ belongs to the optimal policy $\hat{\pi}_{i+1}^{\mathrm{CL}*}(x_{i+1,k})$ without observation delay, the PC agent $i+1$ needs to learn $\hat{\pi}_{i+1}^{\mathrm{CL}*}(x_{i+1,k})$ and the corresponding advantage function in addition to the acting policy $\pi_{i+1}^{\mathrm{CL}}(S^{\rm CL}_{i+1,k})$.}\par

Finally, the following theorem formally justifies the RRA reward function defined above.  
\newtheorem{theorem}{Theorem}
\begin{theorem}
Given the reward function defined in \eqref{comreward_g}, \eqref{comreward_i}, and \eqref{comreward_v}, the expected return as calculated in \eqref{RRA_cumreward} is the same as the optimization objective in \eqref{opti_obj}, where the relationship between the PC reward discount factor and RRA reward discount factor is $\eta=\gamma^{\frac{1}{T}}$. \par
\end{theorem}

The proof of Theorem 1 is given in Appendix B. \par

\subsection{RRA State Definition}	
The state of each RRA agent $i$ is defined as
\begin{align}
S^{\rm CM}_{i,(k,t)}=\{\boldsymbol G_{i,(k,t)},q^{\rm CAM}_{i,(k,t)},\{a^{\rm CL}_{i,k'}\}_{k'=k-\tau_{\mathrm{max}}+1}^{k},t,\epsilon\},
\end{align}
which includes four types of information as discussed below. \par

\paragraph{Channel state information}
$\boldsymbol G_{i,(k,t)}$ is the local observation of CSI at vehicle $i$ in communication interval $(k,t)$, i.e.,
\begin{align}
\boldsymbol G_{i,(k,t)} =&\{G_{i,m,(k,t)},G_{j,i,m,(k,t)},G_{B,i,m,(k,t)},G_{i,B,m,(k,t)}, \IEEEnonumber \\ 
&
G_{m,(k,t)}\}_{m\in\mathcal{M}}.
\end{align}

Specifically, $\boldsymbol G_{i,(k,t)}$ contains the following information: (i) the channel gain of V2V link $i$ over all sub-channels $m\in\mathcal{M}$, $\{G_{ i,m,(k,t)}\}_{m\in\mathcal{M}}$; (ii) the interference channel gain from all V2I link $m$, $\{G_{B,i,m,(k,t)}\}_{m\in\mathcal{M}}$, and other V2V link $j$, $\{G_{j,i,m,(k,t)}\}_{j\ne i,m\in\mathcal{M}}$; (iii) the interference channel gain from V2V link $i$ to the BS over all sub-channel $m\in\mathcal{M}$, $\{G_{i,B,m,(k,t)}\}_{m\in\mathcal{M}}$; (iV) the channel gain of all V2I link $m$, $\{G_{m,(k,t)}\}{m\in\mathcal{M}}$.  Channel gains (i) and (ii) can be accurately estimated by the receiver of V2V link $i$ at the beginning of each communication interval $(k,t)$ \cite{Nasir2018deep}, while (iii) and (iv) are estimated at the BS in each communication interval $(k,t)$ and then broadcast to all vehicles in its coverage, incurring a small signaling overhead. \par

\paragraph{Queue state information}
$S^{\rm CM}_{i,(k,t)}$ includes the queue length $q^{\rm CAM}_{i,(k,t)}$ at vehicle $i$, whose value at communication interval $(k,0)=(k-1,T)$ reflects the observation delay of vehicle $i+1$ at control interval $k$ according to \eqref{eq23}.

\paragraph{PC information}
Since the RRA policy aims at minimizing PC performance degradation due to observation delay, it is our hypothesis that including some PC information in the RRA state could help the RRA agent make more informed decisions. However, it is not a trivial task to determine what information should be included. Ignoring useful information is obviously undesirable, while including useless information should also be avoided due to the curse-of-dimensionality issue. \par  

Based on the following lemmas, Theorem 2 reveals the PC information that should be included in the RRA state. \par 

%	\newtheorem{lemma2}[lemma]{Lemma}
%\begin{lemma2}
%A deterministic PC policy $\pi_{i}^{\mathrm{CL}}(S^{\rm CL}_{i,k})$ based on delayed observation $S^{\rm CL}_{i,k}$ is equivalent to a stochastic policy $\hat{\pi}_{i}^{\mathrm{CL}}(a^{\rm CL}_{i,k}|x_{i,k})$ based on current driving status $x_{i,k}$, where
%	\begin{align}
%	\label{lemma2}
%	\hat{\pi}_{i}^{\mathrm{CL}}(a^{\rm CL}_{i,k}|x_{i,k})=\left\{
%	\begin{array}{ll}
%	\mathrm{Pr}(S^{\rm CL}_{i,k}|x_{i,k}), & \ \mathrm{if} \ a^{\rm CL}_{i,k}=\pi_{i}^{\mathrm{CL}}(S^{\rm CL}_{i,k}) \\
%	0, & \mathrm{Otherwise} \\
%	\end{array}\right. 
%	\end{align}
%\end{lemma2}

\newtheorem{lemma2}[lemma]{Lemma}
\begin{lemma2}
	\label{lemma2}
	The optimal PC policy $\pi_{i}^{\mathrm{CL*}}(S^{\rm CL}_{i,k})$ based on delayed observation $S^{\rm CL}_{i,k}$ is equivalent to the optimal PC policy $\hat{\pi}_{i}^{\mathrm{CL}*}(x_{i,k})$ based on current driving status if $x_{i,k}$ can be fully determined by the delayed observation $S^{\rm CL}_{i,k}$, i.e., $x_{i,k}=f(S^{\rm CL}_{i,k})$ where $f(S^{\rm CL}_{i,k})$ is a deterministic function of $S^{\rm CL}_{i,k}$.
\end{lemma2}

The proof of Lemma 2 is given in {Appendix C}.

\newtheorem{lemma3}[lemma]{Lemma}
\begin{lemma3}
	\label{lemma3}
The current driving status $x_{i,k}$ can be fully determined by the delayed observation $S^{\rm CL}_{i,k}$ if the sequence of actions of vehicle $i-1$ within the time window of $[k-\tau_{i,k},k-1]$ is available to vehicle $i$, i.e., $x_{i,k}=f(S^{\rm CL}_{i,k},\{a^{\rm CL}_{i-1,k'}\}_{k'=k-\tau_{i,k}}^{k-1})$. 
\end{lemma3}

The proof of Lemma 3 is given in {Appendix D}.

\newtheorem{theorem2}[theorem]{Theorem}
\begin{theorem2}
The impact of observation delay $\tau_{i,k}$ to the PC performance of vehicle $i$ depends on the variation of vehicle $i-1$'s control action $a^{\rm CL}_{i-1,k'}$ within the period $k'\in[k-\tau_{\mathrm{max}}+1,k]$.
\end{theorem2}

The proof of Theorem 2 is given in {Appendix E}.

According to Theorem 2, $\{a^{\rm CL}_{i,k'}\}_{k'=k-\tau_{\mathrm{max}}+1}^{k}$ is included in $S^{\rm CM}_{i,(k,t)}$ since it determines how sensitive the PC performance is to the observation delay for a specific control interval $k$. In other words, the characteristics of $a^{\rm CL}_{i-1,k'}$ within the period $k'\in[k-\tau_{\mathrm{max}}+1,k]$ determines the VoI of the received message at vehicle $i$ at control interval $k$. \par

\newtheorem{remark}{Remark}
\begin{remark}[VoI per control interval]
	In the context of NCS or task-oriented scheduling, Age of Information (AoI) and VoI are two important metrics. {AoI captures the importance of information by measuring its timeliness attribute \cite{kaul2012real}, while VoI measures how much the recipient of the information can reduce the uncertainty of the stochastic processes related to decision-making \cite{kosta2017age,wang2020value,wang2022framework}.} It is shown in \cite{ayan2019age} that VoI is related to AoI while providing more accurate guidance to the scheduler design. However, VoI is normally {derived based on mutual information} as a single value for the whole control task, ignoring the variation in the significance of transmitted data during the control episode. In this paper, we {design the RRA reward function to capture VoI based on the PC performance degradation due to the observation delay of shared information in each control interval.} Moreover, we include the PC information in the RRA state, so that the RRA decisions are made based on the \emph{VoI per control interval} that captures VoI at a finer time granularity.  
\end{remark}   

\paragraph{Other information}

In order to stabilize experience replay for multi-agent DRL, we augment $S^{\rm CM}_{i,(k,t)}$ with the exploration rate $\epsilon$ \cite{foerster2017stabilising}, i.e., the probability of random selecting action in $\epsilon$-greedy policy for Deep Q Networks (DQN).\par

Finally, the index of communication interval $t$ is a part of the RRA state since it relates to the reward function in \eqref{comreward_g}. \par

\subsection{RRA Action}	

The RRA action of each vehicle $i\in\mathcal{V}\backslash\{N-1\}$ at communication interval $(k,t)$ includes the decisions on the sub-channel allocation $\{\theta_{i,m,(k,t)}\}_{m\in\mathcal{M}}$ and transmit power $\{P^{\rm V}_{i,m,(k,t)}\}_{m\in\mathcal{M}}$ for V2V link $i$. Since we consider that at most one sub-channel is allocated to a V2V link at any communication interval $(k,t)$, i.e., $\sum_{m=0}^{M-1} \theta_{i,m,(k,t)} \leq 1$, we can simplify the sub-channel allocation action from $\{\theta_{i,m,(k,t)}\}_{m\in\mathcal{M}}$ to $m_{i}\in\mathcal{M}\cup\{-1\}$, where $m_{i}\in \mathcal{M}$ means that sub-channel $m_{i}$ is selected and $m_{i}=-1$ means that no sub-channel is allocated to V2V link $i$.\par

We discretize the transmit power to four levels for ease of learning and practical circuit restriction \cite{liang2019spectrum,xiang2021multi}, i.e., $P^{\rm V}_{i,m,(k,t)}\in\mathcal{A}_P=\{23, 15, 5, -100\}$ $\rm dBm$. Noted that $-100$ dBm can be considered as zero transmit power. \par

Therefore, the RRA action of each vehicle $i\in\mathcal{V}\backslash\{N-1\}$ at communication interval $(k,t)$ is defined as
\begin{align}\label{CV2Xaction}
a^{\rm CM}_{i,(k,t)}=\left\{m_{i},P^{\rm V}_{i,m,(k,t)}| m_{i}\in \mathcal{M}\cup\{-1\}, P^{\rm V}_{i,m,(k,t)} \in \mathcal{A}_P\right\}.
\end{align}\par

The objective of the RRA problem is to find the optimal policy $\pi^{\rm CM*}$ that maximizes the expected return $J^{\rm CM}$, i.e.,
\begin{equation}
		\pi^{\rm CM*}=\arg\max_{\boldsymbol{\pi^{\rm CM}}}J^{\rm CM}.
\end{equation}  

\section{DRL Solution for RRA}
The proposed DRL algorithm to solve the RRA problem is based on Double Deep Q Networks (DDQN)\cite{van2016deep}, which is an improved version of the classical DQN algorithm, and has been widely adopted to deal with RL tasks with discrete action space. DQN combines Q-learning with a deep neural network (DNN) $Q^{\rm CM}_{i}(S^{\rm CM}_{i,(k,t)}, a^{\rm CM}_{i,(k,t)} |\theta_i)$, which acts as a function approximator and maps states to actions’ value estimation\cite{mnih2015human}. Since DQN's max operator selects and evaluates action $a^{\rm CM}_{i,(k,t)}$ using the same value, the overestimated value is more likely to be chosen, leading to over-optimistic estimates. In order to avoid this, DDQN is proposed for decoupling the selection of actions and their evaluation in the target maximization\cite{van2016deep}. Specifically, the action $a^{\rm CM}_{i,(k,t)}$ is chosen by the current Q networks $Q^{\rm CM}_{i}(S^{\rm CM}_{i,(k,t)}, a^{\rm CM}_{i,(k,t)} |\theta_i)$ with the maximum Q value, and the evaluation of the action $a^{\rm CM}_{i,(k,t)}$ is used by the target Q network $Q^{\rm CM'}_{i}(S^{\rm CM}_{i,(k,t)}, a^{\rm CM}_{i,(k,t)} |\theta'_i)$.\par

When tackling the RRA problem in this paper, an effective policy cannot be learned by direct application of DDQN due to the following two challenges.

\begin{description}
	\item[Multi-agent problem] Similar to PC, RRA is also a multi-agent problem. However, the simple IL approach is not suitable for RRA since the agents cooperatively optimize the \emph{global} expected cumulative reward, and it is hard for each agent to deduce its own contribution to the global reward. The challenge in distinguishing the agents' credit makes the learning of an effective policy non-trivial.
	\item[Long-range planning with sparse reward problem] According to \eqref{opti_obj}, the RRA problem involves optimization over a time horizon of $KT$ communication intervals, where $T$ consecutive communication intervals correspond to a control interval. In the reward function defined in \eqref{comreward_g}, $R_{{\rm I},(k,t)}$ is provided to the agent in every communication interval $(k,t)$, while $R_{{\rm V},(k,T)}$ is only available at the last communication interval $T-1$ of each control interval $k\in\mathcal{K}$. The sparse-reward tasks coupled with long-horizon planning become prohibitively difficult as highly specific action sequences must be executed prior to observing any nontrivial feedback. 
\end{description}

In the following, we propose the MTCC-RRA algorithm based on DDQN, where we apply the \emph{reward shaping} and \emph{RBPER}\cite{zhong2017reward} techniques to deal with these two challenges, respectively.\par

\subsubsection{Reward shaping}
To overcome the multi-agent challenge as described above, we design an individual RRA reward $R^{\rm CM}_{i,(k,t)}$ for each vehicle $i\in\mathcal{V}\backslash\{N-1\}$ as  
\begin{align}
& \label{comreward}
R^{\rm CM}_{i,(k,t)}= \\ \IEEEnonumber
&\left\{
\begin{array}{ll}
\kappa_1 \hat{R}_{{\rm I},i,(k,t)}, &   t< T-1\\
\kappa_1 \hat{R}_{{\rm I},i,(k,t)}+\\ \kappa_2 A_{\hat{\pi}^{\rm CL*}_{i+1}}\left(x_{i+1,k+1},\pi^{\rm CL}_{i+1}(S^{\rm CL}_{i+1,k+1})\right) , & \  t= T-1 \\
\end{array}\right. ,
\end{align}
\noindent where
\begin{align}
\label{diff_reward}
\hat{R}_{{\rm I},i,(k,t)} =r_{m,(k,t)}-r_{m,(k,t)}^{\bar{i}} 
\end{align}
\noindent is the difference reward that reflects the data rate loss of V2I link $m$ resulting from sharing the sub-channel with V2V link $i$, while the sub-channel allocations of the other V2V links remain the same. $r_{m,(k,t)}$ is the actual data rate of V2I link $m$ at communication interval $(k,t)$, while $r_{m,(k,t)}^{\bar{i}}$ is the data rate of V2I link $m$ if V2V link $i$ does not occupy sub-channel $m$. All the channel gains for the calculation of $r_{m,(k,t)}^{\bar{i}}$ are included in the state space. Since $\hat{R}_{{\rm I},i,(k,t)}$ is a negative value, maximizing its value over time is equivalent to minimizing the throughput loss of V2I links caused by interference from V2V link $i$. \par

Moreover, since the RRA decisions at vehicle $i$ will only affect the observation delay and thus the PC performance of its follower $i+1$, the advantage function $A_{\hat{\pi}^{\rm CL*}_{i+1}}\left(x_{i+1,k+1},\pi^{\rm CL}_{i+1}(S^{\rm CL}_{i+1,k+1})\right)$ is a part of the individual reward of vehicle $i$ at each communication interval $(k,T-1)$ in \eqref{comreward}.\par

{
 \newtheorem{remark2}[remark]{Remark}
 \begin{remark2}[Obtaining $\hat{R}_{{\rm I},i,(k,t)}$ and $A_{\hat{\pi}^{\rm CL*}_{i+1}}$ in \eqref{comreward}]
We consider the centralized training decentralized execution paradigm \cite{lowe2017multi} for MTCC-RRA, so that the global state and global action are available to the RRA agents during training. Therefore, both $r_{m,(k,t)}$ and $r_{m,(k,t)}^{\bar{i}}$ can be calculated by \eqref{SINRV2I} and \eqref{rateV2I}, where $\theta_{i,m,(k,t)}$ in \eqref{SINRV2I} for V2V link $i$ is set to $1$ and $0$, respectively. Then, $\hat{R}_{{\rm I},i,(k,t)}$ can be derived by \eqref{diff_reward}. Moreover, the PC advantage function $A_{\hat{\pi}^{\rm CL*}_{i+1}}\left(x_{i+1,k+1},\pi^{\rm CL}_{i+1}(S^{\rm CL}_{i+1,k+1})\right)$ of follower $i+1$ is available to its predecessor $i$ due to the centralized training assumption. {Note that each RRA agent trains a local critic instead of all the agents collectively training a centralized critic, since the global reward is decomposed into the individual rewards of the RRA agents.}	  
 \end{remark2}  }

\subsubsection{Reward backpropagation prioritized experience replay}

To resolve the difficulty of training for long-range planning with sparse reward, RBPER assigns a higher sampling priority $\beta$ to the transition $\left(S^{\rm CM}_{i,(k,T-1)}, a^{\rm CM}_{i,(k,T-1)}, R^{\rm CM}_{i,(k,T-1)}, S^{\rm CM}_{i,(k,T)}\right)$ of the last communication interval $T-1$ of each control interval $k$ and then propagates the priority $\beta$ back to the transition of the previous communication interval $T-2$ once it has been sampled for training and so on. The process will keep going until $\beta$ is propagated to the transition $\left(S^{\rm CM}_{i,(k,0)}, a^{\rm CM}_{i,(k,0)}, R^{\rm CM}_{i,(k,0)}, S^{\rm CM}_{i,(k,1)}\right)$ of the first communication interval $0$ of each control interval $k$. Then priority $\beta$ will be propagated back to the transition $\left(S^{\rm CM}_{i,(k,T-1)}, a^{\rm CM}_{i,(k,T-1)}, R^{\rm CM}_{i,(k,T-1)}, S^{\rm CM}_{i,(k,T)}\right)$ and start a new round of propagation. The high priority will be decayed by factor $\zeta \in (0,1)$, once we start a new round until it goes down to normal priority $1$. Thus, the target Q, $Q_i^{\rm CM'}(S^{\rm CM}_{i,(k,t+1)}, a^{\rm CM}_{i,(k,t+1)} |\theta'_i)$, of the previous communication interval $t$ will be established on the basis of the trained Q, $Q_i^{\rm CM}(S^{\rm CM}_{i,(k,t+1)}, a^{\rm CM}_{i,(k,t+1)} |\theta_i)$, of the next communication interval $t+1$, improving the accuracy of Q value estimation and training efficiency.\par

\section{Joint Optimization of Multi-timescale Control and Communications}
As mentioned above, it is assumed that the RRA (resp. PC) policy is available when training the PC (resp. RRA) policy. Due to the interactions between RRA and PC, it is not possible to learn the optimal RRA or PC policy first without knowing the other optimal policy. To solve this dilemma, the most straightforward approach is to train the proposed MTCC-PC and MTCC-RRA algorithms simultaneously. However, training both algorithms from scratch poses a great challenge in convergence, especially for RRA whose rewards depend on the PC. The rewards calculated from a PC model that has not reached convergence yet are misleading to RRA, resulting in substantial difficulty in learning an efficient RRA policy. Moreover, it is more challenging for {MTCC-RRA algorithm} to reach convergence than {MTCC-PC algorithm} under the multi-timescale framework, since the number of communication intervals is $T$ times that of the control intervals per episode. In this paper, we propose to iteratively train {MTCC-PC and MTCC-RRA algorithms}, where each iteration $z\in\{1,\cdots,Z\}$ consists of the following two steps.  

\begin{description}
\item[Step 1] The {MTCC-PC} is trained for $E^{\rm CL}$ episodes under the observation delay induced by the RRA policy learned in Step 2 of iteration $z-1$. The RRA transitions $\left(S^{\rm CM}_{i,(k,t)}, a^{\rm CM}_{i,(k,t)}, R^{\rm CM}_{i,(k,t)}, S^{\rm CM}_{i,(k,t+1)}\right)$ are stored in an experience replay buffer $D_i^{\rm CM}$ for all episodes $E>E^{\rm CL_{Th}}$.
\item[Step 2] The{ MTCC-RRA} is trained for $E^{\rm CM}$ episodes, where the PC model learned in Step 1 of iteration $z$ is used to calculate vehicle trajectories and RRA rewards. The PC transitions $\left(S^{\rm CL}_{i,k}, a^{\rm CL}_{i,k}, R^{\rm CL}_{i,k}, S^{\rm CL}_{i,k}\right)$ are stored in an experience replay buffer $D_i^{\rm CL}$ for all episodes $E>E^{\rm CM_{Th}}$.
%\item[Step 3] Training round $2$ for fewer training episodes are conducted where Step 1 and Step 2 in Training round $1$ are repeated, thus the trained PC and RRA models are further trained to improve performance. 
\end{description} 

%which is not computationally efficient since the PC or RRA policy is frozen when learning the other policy, and it might take multiple iterations to obtain satisfactory results. To overcome the above challenges, we propose a joint training approach as below.\par

Note that it is essential to train the {MTCC-PC} first in Step 1, and the {MTCC-RRA} next in Step 2. This is because RRA training heavily depends on the PC advantage function to calculate reasonable reward signals as given in \eqref{comreward_v}. The initial RRA policy randomly selects the RRA actions according to uniform distribution.\par

To improve the sample efficiency, the RRA (resp. PC) experience is stored in the experience replay buffer $D_i^{\rm CM}$ (resp. $D_i^{\rm CL}$) when training the PC (resp. RRA) model. The stored experience can be leveraged at the initial phase of training when RRA (resp. PC) is trained in Step 2 (resp. Step 1 of the next iteration). To ensure that the experience are generated by relatively well-learned PC (resp. RRA) policies in the previous step, only the transitions at the later stage of PC (resp. RRA) training when $E>E^{\rm CL_{Th}}$ (resp. $E>E^{\rm CM_{Th}})$ are used. \par

Although it takes multiple iterations to obtain optimal results in MTCC, we will demonstrate in Section VI that satisfactory policies can be learned in only one iteration.\par

The proposed MTCC training algorithm is given in Algorithm \ref{alg1}, {where the MTCC-PC and MTCC-RRA algorithms are iteratively trained for $Z$ times. In each iteration, the function $\mathrm{MTCC}(\mathrm{PC\_TRAIN=1}, \mathrm{RRA\_TRAIN=0}, E^{\rm CL}, \mathrm{ITER=}$\\$z)$ corresponds to the MTCC-PC algorithm and the function $\mathrm{MTCC}(\mathrm{PC\_TRAIN=0}, \mathrm{RRA\_TRAIN=1}, E^{\rm CM}, \mathrm{ITER=}$\\$z)$ corresponds to the MTCC-RRA algorithm. To improve the sample efficiency, the transitions at the later stage of PC (resp. RRA) training when $E>E^{\rm CL_{Th}}$ (resp. $E>E^{\rm CM_{Th}})$ are used at the initial phase of training when RRA (resp. PC) is trained in Step 2 (resp. Step 1 of the next iteration). The pseudocode of the function MTCC is provided in Funtion \ref{function2}. When the input argument $\mathrm{PC\_TRAIN}=1$, the DDPG algorithm is used to train the PC agent. When the input argument $\mathrm{RRA\_TRAIN}=1$, the DDQN-RBPER algorithm is used to train the RRA agent.} The pseudocode of the DDPG algorithm and DQN-RBPER algorithm can be found in \cite{lillicrap2015continuous} and \cite{zhong2017reward}, respectively. The DDQN-RBPER algorithm extends DQN-RBPER with DDQN \cite{van2016deep}.  \par

%{Note that before training the MTCC algorithm, we will first learn the optimal PC policy $\hat{\pi}_{i}^{\mathrm{CL}*}(x_{i,k})$ under the undelayed driving status $x_{i,k}$ to get the undelayed Q-value for the calculation of the advantage function during MTCC training.}\par

\begin{figure}[tpb!]
\centering
\includegraphics[width=0.45\textwidth]{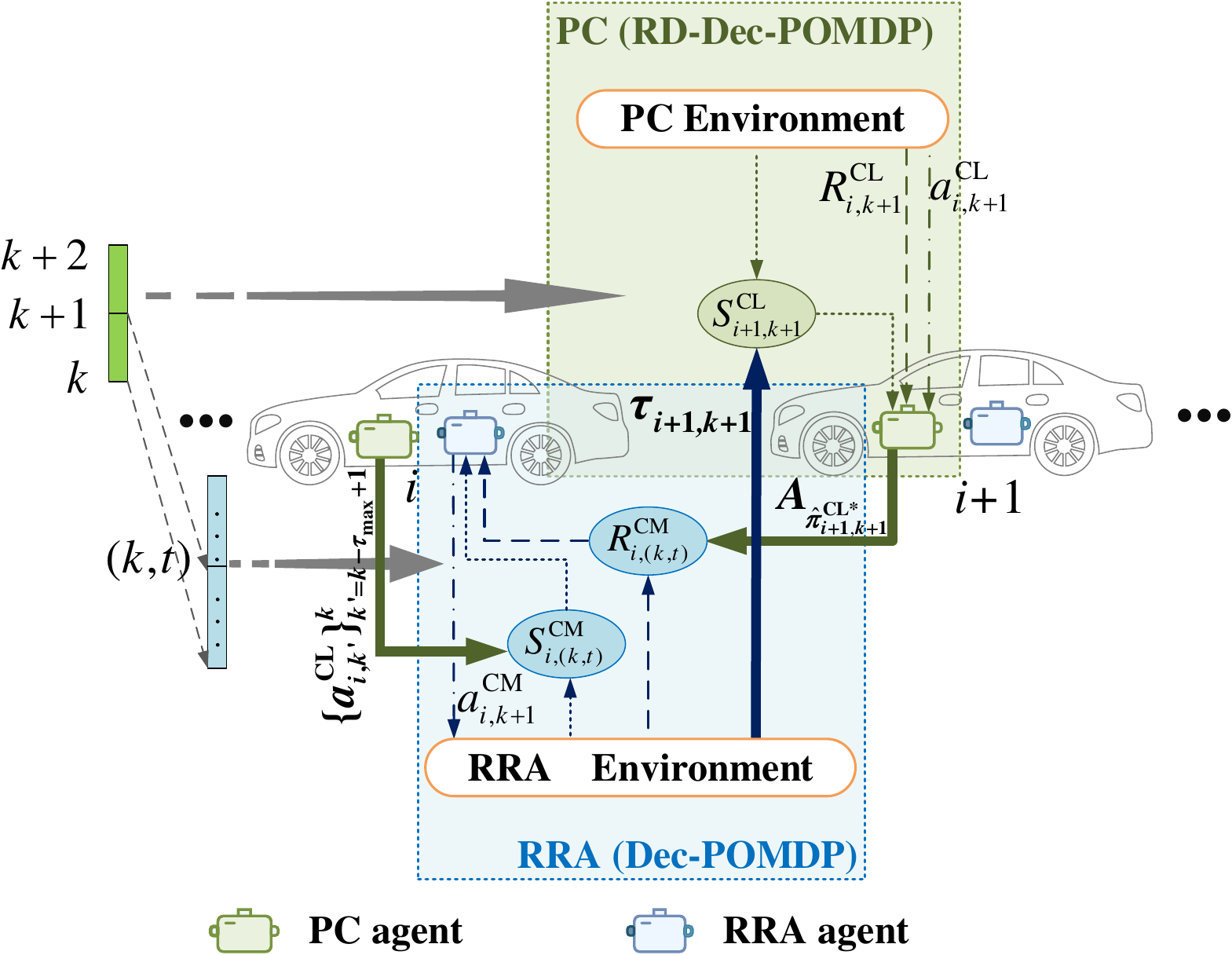}
\caption{The framework of MTCC algorithm}
\label{MTCC_framework}
\end{figure}

The unified DRL training framework for MTCC is illustrated in Fig.~\ref{MTCC_framework}. Each vehicle $i$ contains a PC agent and an RRA agent, except for the leading vehicle $0$ {which} only has an RRA agent and the last vehicle $N-1$ which only has a PC agent. The environment can also be divided into the PC environment, which provides driving status and PC rewards to the PC agents; and the RRA environment, which provides wireless channel and queue states as well as V2I-related rewards to the RRA agents. Meanwhile, the PC (resp. RRA) agents also interact with the RRA (resp. PC) environment/agents. For each RRA agent of vehicle $i\in\mathcal{V}\backslash \{N-1\}$, the PC advantage function $A_{\hat{\pi}^{\rm CL*}_{i+1,k+1}}$ of vehicle $i+1$ at control interval $k+1$ is used to calculate the PC-related portion of RRA reward ${R}^{\rm CM}_{i,(k,T-1)}$ at communication interval $(k,T-1)$. Moreover, the RRA state ${S}^{\rm CM}_{i,(k,t)}$ of vehicle $i$ at communication interval $(k,t)$ is augmented with its PC action history $\{a^{\rm CL}_{i,k'}\}_{k'=k-\tau_{\mathrm{max}}+1}^{k}$. Meanwhile, for each PC agent of vehicle $i+1\in\mathcal{V}\backslash \{0\}$, its state ${S}^{\rm CL}_{i+1,k+1}$ at control interval $k+1$ depends on the observation delay $\tau_{i+1,k+1}$, which is provided by the RRA environment and determined by the queue state $q_{i,(k,T)}^{\mathrm{CAM}}$ of vehicle $i$ at communication interval $(k,T)$.  \par

{After the PC agents and RRA agents are trained in the centralized fashion, each vehicle makes PC and/or RRA decisions in a decentralized manner based on its local observations of the PC and RRA states. The decisions can be made in real-time since only the forward propagation in the DNNs is involved during execution. }\par
  
\floatname{algorithm}{Algorithm} 
\setcounter{algorithm}{0}
\begin{algorithm}[h]
	\caption{MTCC algorithm}
	\label{alg1}
	\begin{algorithmic}[1]
		\FOR {follower $i\in\mathcal{V}\backslash \{0\}$	}
         \STATE Initialize PC actor network $\mu^{\rm CL}_i(s|\theta^\mu_i)$ and critic network $Q^{\rm CL}_i(s,u|\theta^Q_i)$ with random weights $\theta^\mu_i$ and $\theta^Q_i$
         \STATE Initialize PC target network weights with $\theta^{\mu'}_i=\theta^{\mu}_i$ and $\theta^{Q'}_i=\theta^{Q}_i$
         \STATE Initialize replay buffer $D_{i}^{\rm CL}$
         \ENDFOR
         \FOR {predecessor $i\in\mathcal{V}\backslash \{N-1\}$ }
         \STATE Initialize RRA Q networks $Q^{\rm CM}_{i}(s,a|\theta_{i})$ with random weights  $\theta_{i}$
         \STATE Initialize RRA target network weights with $\theta'_{i}=\theta_{i}$
         \STATE Initialize replay buffer $D_{i}^{\rm CM}$ 
          \ENDFOR		
        \FOR {Iteration $z=1,...,Z$ }
        \STATE \textbf{Step 1}: 
        \STATE Function $\mathrm{MTCC}(\mathrm{PC\_TRAIN=1}, \mathrm{RRA\_TRAIN=0},$\\ $E^{\rm CL}, \mathrm{ITER=}z)$ 
        \STATE Delete RRA experience for episodes $E\leq E^{\rm CL_{Th}}$ from $D_{i}^{\rm CM}$
        \STATE \textbf{Step 2}: 
        \STATE Function $\mathrm{MTCC}(\mathrm{PC\_TRAIN=0}, \mathrm{RRA\_TRAIN=1},$\\ $E^{\rm CM}, \mathrm{ITER=}z)$ 
        \STATE Delete PC experience for episodes $E\leq E^{\rm CM_{Th}}$ from $D_{i}^{\rm CL}$
        \ENDFOR
%       \STATE \textbf{Step 3}: 
%       \STATE Repeat Step 1 and Step 2 with trained PC and RRA for fewer episodes to conduct Training rounds $2$, where $\epsilon=0$.
    \end{algorithmic}
\end{algorithm}

\floatname{algorithm}{Function} 
\setcounter{algorithm}{0}
\begin{algorithm}[h]
	\renewcommand{\algorithmicrequire}{\textbf{Input:}}
	\renewcommand{\algorithmicensure}{\textbf{Output:}}
	\caption{$\mathrm{MTCC}(\mathrm{PC\_TRAIN}, \mathrm{RRA\_TRAIN}, E, \mathrm{ITER})$}
	\label{function2}
	\begin{minipage}{\columnwidth}
		\begin{algorithmic}[1]	
%  		    \IF{$\mathrm{RRA\_TRAIN}\neq 1$}
%                \STATE Set initial exploration rate $\epsilon=0$
%			\ENDIF
%			\IF{$\mathrm{ITER}= 1$}
%			\STATE Set initial exploration rate $\epsilon=1$
%			\ENDIF
			\FOR{episode $e = 1,\dots, E$ }
			\FOR{control interval $k = 0,\dots, K-1$ }
			\FOR{communication interval $t=0,\dots, T-1$}
			\FOR{predecessor $i\in\mathcal{V}\backslash \{N-1\}$ }
			\STATE Update the vehicle position and receive RRA state $S^{\rm CM}_{i,(k,t)}$ 
            \IF{$\mathrm{RRA\_TRAIN}= 1$}
            \STATE Select RRA action $a^{\rm CM}_{i,(k,t)}$ according to the $\epsilon$-greedy policy with respect to $Q^{\rm CM}_{i}(S^{\rm CM}_{i,(k,t)},a|\theta_{i})$
            \ELSE
            \IF{$\mathrm{ITER}= 1$ }
            \STATE Select a random RRA action $a^{\rm CM}_{i,(k,t)}$ 
            \ELSE
            \STATE Select RRA action $a^{\rm CM}_{i,(k,t)}$ according to the greedy policy with respect to $Q^{\rm CM}_{i}(S^{\rm CM}_{i,(k,t)},a|\theta_{i})$
            \ENDIF
            \ENDIF
%            $\mathrm{ITER}= 1$ and 
%                
%                \IF{$\mathrm{ITER}= 1$ and $\mathrm{RRA\_TRAIN}= 0$}
%                \STATE Select a random RRA action $a^{\rm CM}_{i,(k,t)}$ 
%                \ELSE
%			\STATE With probability $\epsilon$ select a random RRA action $a^{\rm CM}_{i,(k,t)}$
%			\STATE otherwise select RRA action $a^{\rm CM}_{i,(k,t)}=\arg\max_{a}Q^{\rm CM}_{i-1}(S^{\rm CM}_{i,(k,t)},a|\theta_{i})$
%                \ENDIF
			\STATE Execute RRA action $a^{\rm CM}_{i,(k,t)}$ and observe RRA reward ${R}^{\rm CM}_{i,(k,t)}$ and next RRA state $S^{\rm CM}_{i,(k,t+1)}$
			\STATE Store RRA transition $(S^{\rm CM}_{i,(k,t)}, a^{\rm CM}_{i,(k,t)},R^{\rm CM}_{i,(k,t)},$ \\$S^{\rm CM}_{i,(k,t+1)})$ with priority $p$ in $D_i^{\rm CM}$
			\IF{$\mathrm{RRA\_TRAIN}=1$}
			\STATE Update $\theta_i,\theta'_i$ with DDQN-RBPER algorithm
			\ENDIF
			\ENDFOR
			\ENDFOR
			\FOR{follower $i\in\mathcal{V}\backslash \{0\}$}
			\STATE{Calculate observation delay $\tau_{i,k}$ and receive PC state $S^{\rm CL}_{i,k}$}
			\STATE Select PC action $a^{\rm CL}_{i,k}=\mu^{\rm CL}_i(S^{\rm CL}_{i,k}|\theta^\mu_i)$ 
		    \IF{$\mathrm{PC\_TRAIN}=1$}
			\STATE Add exploration noise to PC action $a^{\rm CL}_{i,k}=a^{\rm CL}_{i,k}+\mathcal{N}_t$
			\ENDIF
			\STATE Store PC transition $(S^{\rm CL}_{i,k},a^{\rm CL}_{i,k}, R^{\rm CL}_{i,k}, S^{\rm CL}_{i,k+1})$ in $D_i^{\rm CL}$
		    \IF{$\mathrm{PC\_TRAIN}=1$}
			\STATE Update $\theta^\mu_i,\theta^Q_i,\theta^{\mu'}_i,\theta^{Q'}_i$ with DDPG algorithm
			\ENDIF
			\ENDFOR
			\ENDFOR
			\ENDFOR
%			\IF{$ts=1$}
%			\RETURN $\{\theta^\mu_i,\theta^Q_i,D_{i-1}^{\rm CM} \}_{i=1}^{N-1}$
%			\ELSIF{$ts=2$}
%			\RETURN $\{\theta_{i-1} \}_{i=1}^{N-1} $
%			\ELSE
%			\RETURN $\{\theta^\mu_i,\theta^Q_i, \theta_{i-1} \}_{i=1}^{N-1}$
%			\ENDIF
		\end{algorithmic}
	\end{minipage}
\end{algorithm}

\section{Experimental Results}
In this section, we design experiments to answer the following questions: (1) Can MTCC-RRA outperform the state-of-the-art control-aware communications by capturing ``VoI per control interval" through its state space and reward design? (2) Can the reward shaping and RBPER mechanisms efficiently tackle the multi-agent and sparse reward problems for MTCC-RRA? (3) Can the satisfactory policies for MTCC be learned in only one iteration? The experimental results of the proposed MTCC algorithm and the baseline DRL algorithms are presented and compared for the above purposes. \par

\subsection{Experimental Setup}
We conduct experiments based on the PC and RRA environments provided in Part I of this { two-part} paper. In addition to the hyper-parameters of DDPG for solving the communication-aware PC problem, we summarized the hyper-parameters for DDQN solving the control-aware RRA problem in Table \ref{hyper_parameters}. The values of all the hyper-parameters were selected by performing a grid search as in \cite{mnih2015human}, using the values reported in\cite{van2016deep} as a reference. DDQN has two hidden layers with $256$ and $64$ nodes, respectively, where $128$ of the nodes in the first hidden layer are modeled as Long Short-Term Memory (LSTM) for processing PC action history $\{a^{\rm CL}_{i,k'}\}_{k'=k-\tau_{\mathrm{max}}+1}^{k}$ in $S^{\rm CM}_{i,(k,t)}$ and the rest of the 128 nodes in the first hidden layer are used for the remaining dimensions of state $S^{\rm CM}_{i,(k,t)}$. Then all the $256$ nodes are connected to the second layer with $64$ nodes. The target network of DDQN is updated every $4$ communication intervals.
\begin{table}[t]
	\renewcommand{\arraystretch}{1.3}
	\caption{Hyper-Parameters of the DRL algorithms for training} \label{alg_para} \centering
	\begin{tabular}{ll}
		\hline
		{\textbf{Parameter}} & \textbf{Value} \\
		\hline
            Network size &$256,64$ \\
		\hline
            Activation function  &relu, relu, relu, linear \\
            \hline
            Learning rate & {$0.0001$}\\
            \hline
             Exploration rate $\epsilon$ & $1 \rightarrow 0.05$\\
            \hline
            Higher sampling priority $\beta$  &$100$ \\
            \hline
            Decay factor $\zeta$  &$0.2$ \\    
		\hline
		Batch size $N_b$ &$64$ \\
		\hline
		Replay buffer size &$200000$ \\
		\hline
		Reward discount factor $\gamma$ & 0.99979\\
		\hline
            \makecell[l]{Target update frequency\\ of DDQN $N^-$}& $4$    \\
		\hline
		\makecell[l]{Final layer \\weights/biases initialization} & \makecell[l]{Random uniform distribution \\$[-3\times10^{-3},3\times10^{-3}]$} \\
		\hline
		\makecell[l]{Other layer \\weights/biases initialization} &\makecell[l]{Random uniform distribution$[-\frac{1}{\sqrt{f}},\frac{1}{\sqrt{f}}]$\\($f$ is the fan-in of the layer)} \\
		\hline
		\label{hyper_parameters}
	\end{tabular}
\end{table}
\subsubsection{Baseline algorithms}

\begin{itemize}
	\item  \textbf{Delay-aware RRA (Delay-RRA)}, which is an improved version of the DRL-based RRA algorithm for C-V2X communications in \cite{liang2019spectrum,xiang2021multi,zhang2022mean}. \cite{liang2019spectrum} focuses on RRA within one control interval and aims to minimize the latency and maximize the success probability of delivering the CAM generated at the beginning of the control interval. To deal with the sparse reward problem, the V2V transmission rate at each time step is included in the reward function to expedite convergence. Since \cite{liang2019spectrum} does not address the credit assignment problem in MARL, we improve the algorithm by applying reward shaping similar to MTCC for a fair comparison. Moreover, to apply the algorithm in a longer time horizon with multiple control intervals, we assume that an agent replaces any old CAM that has not yet been fully delivered in the previous control interval with the newly generated CAM at the beginning of each control interval. Therefore, the reward function $R^{\rm IR}_{i,(k,t)}$ of Delay-RRA is given as 
	\begin{align}
	\label{IR}
	R^{\rm IR}_{i,(k,t)}=\lambda_1 \hat{R}_{{\rm I},i,(k,t)}+\lambda_2 (r_{i,(k,t)}|_{q^{\rm CAM}_{i,(k,t)}>0}+G|_{q^{\rm CAM}_{i,(k,t)}=0}),
	\end{align}
	\noindent where $\lambda_1=0.001/W$ and $\lambda_2=0.1/W$. Notice that $G$ is a tuned hyper-parameter encouraging early completion of transmission for CAM whose value is greater than the largest $r_{i,(k,t)}$ ever obtained. We set $G=10W$ in the experiments.
    \item \textbf{AoI-aware RRA (AoI-RRA)}, which is similar to the DRL-based RRA in MTCC except that it aims to minimize cumulative AoI instead of maximizing cumulative ``VoI per control interval". Specifically, the advantage function $A_{\hat{\pi}^{\rm CL*}_{i+1}}\left(x_{i+1,k+1},\pi^{\rm CL}_{i+1}(S^{\rm CL}_{i+1,k+1})\right)$ in \eqref{comreward} is replaced by $-AoI_{i+1,k}$, where $AoI_{i+1,k}$ is the AoI of vehicle $i+1$ that evolves according to  
    \begin{align}
 AoI_{i+1,k+1}= 
 \left\{
 \begin{array}{ll}
 T & \mathrm{if} \ q^{\rm CAM}_{i,(k,T)}=0  \\
 AoI_{i+1,k}+T & \mathrm{if} \ q^{\rm CAM}_{i,(k,T)}>0  \\
 \end{array}\right. .
 \end{align}
For Delay-RRA and AoI-RRA, $q^{\rm CAM}_{i,(k,0)}$ is always $1$ and no longer equals to $q^{\rm CAM}_{i,(k-1,T)}$. \par
\item \textbf{MTCC without reward shaping (MTCC\_wo\_RS)}, where the global reward in \eqref{comreward_g} is used in MTCC for training DRL-based RRA, instead of the individual reward in \eqref{comreward} after reward shaping.
\item \textbf{MTCC without RBPER (MTCC\_wo\_RBPER)}, where DDQN without RBPER is used in MTCC for training DRL-based RRA.
\end{itemize}

\subsubsection{Experiment design}
To answer the three questions mentioned at the beginning of Section VI, we design the following three experiments. The proposed MTCC is trained in one iteration for Experiments (1) and (2).
\begin{itemize}
    \item \textbf{Experiment (1): Performance comparison of MTCC-RRA with Delay-RRA and AoI-RRA.} To answer Question (1), all three algorithms are trained for $E^{\rm CM}=100$ episodes. The trained PC policy by MTCC-PC is adopted during both the training and testing of the RRA algorithms. Therefore, the PC performance difference between Delay-RRA, AoI-RRA, and MTCC-RRA is only due to the difference in induced observation delay by the RRA algorithms.
     \item \textbf{Experiment (2): Performance comparison of MTCC-RRA with MTCC\_wo\_RS and MTCC\_wo\_RBPER.} To answer Question (2), MTCC\_wo\_RS and MTCC\_wo\_RBPER are trained for $E^{\rm CM}=100$ episodes and the resulting performance is compared with that of MTCC-RRA.
     \item \textbf{Experiment (3): Performance Comparison of MTCC in one and two iterations.} To answer Question (3), we train MTCC for a second iteration and observe the resultant performance gain.
\end{itemize}

Note that the episode thresholds for storing experience are set to $E^{\rm CL_{Th}}=\frac{E^{\rm CL}}{5}$ and  $E^{\rm CM_{Th}}=\frac{E^{\rm CM}}{5}$, respectively.  \par 

\subsection{Performance Comparison of MTCC-RRA, Delay-RRA, and AoI-RRA.}
\subsubsection{Performance for testing data}

\begin{table*}[htb!]
	\centering
	\caption{Performance after training RRA in one iteration. We present RRA performance and average sum V2I throughput, and sum PC performance for MTCC-RRA, Delay-RRA, and AoI-RRA, MTCC\_wo\_RS, and MTCC\_wo\_RBPER, respectively.}
	\begin{tabular}{|c|ccccc|}
		\hline
		& \textbf{MTCC-RRA}&\textbf{Delay-RRA}&\textbf{AoI-RRA}&\textbf{MTCC\_wo\_RS}&\textbf{MTCC\_wo\_RBPER}\\
		\hline
		\textbf{RRA performance}&68.2553&43.0803&43.8063&58.3390&-209.1657 \\
		\hline
		\textbf{Average sum V2I throughput} ($\rm {Mbps}$)&4.3982&3.6646&3.6432&4.0040&4.7523\\
  		\hline
		\textbf{Sum PC performance}&-1.5390&-1.4231&-1.4231 &-3.2671&-28.1860\\
		\hline
	\end{tabular}
	\label{table_Performance_Step2}
\end{table*}
The RRA performance are reported in Table~\ref{table_Performance_Step2} for MTCC-RRA, Delay-RRA, and AoI-RRA, respectively, which are obtained by averaging the RRA return over $100$ testing episodes when training is completed. The RRA return is derived as the cumulative global RRA reward defined in \eqref{comreward_g} overall communication intervals in one control episode. Since the global RRA reward consists of two parts, i.e., the sum throughput of all V2I links and the sum PC performance degradation due to delayed observation for all followers, we also report the average sum V2I throughput and sum PC performance in Table~\ref{table_Performance_Step2}. The former is obtained by averaging the sum throughput for all V2I links over $100$ testing episodes, while the latter is obtained by averaging the sum PC returns of all followers. Note that the sum PC performance and the sum PC performance degradation are essentially the same since their values are different by the same fixed amount for all the algorithms, where the difference corresponds to the optimal sum PC performance without observation delay. \par

As shown in Table~\ref{table_Performance_Step2}, MTCC-RRA outperforms AoI-RRA and Delay-RRA algorithms by $58.44\%$ and $55.81\%$, respectively, in terms of RRA performance. The result is not surprising since MTCC-RRA aims at maximizing the RRA performance corresponding to the cumulative global RRA reward in \eqref{comreward_g}, while Delay-RRA and AoI-RRA have different reward functions. \par   

The reward function of MTCC-RRA differs from those of Delay-RRA and AoI-RRA in that MTCC-RRA aims at minimizing the PC performance degradation due to observation delay instead of minimizing the delay or AoI, while all the reward functions also target at maximizing the V2I throughput. Therefore, MTCC-RRA can achieve larger V2I throughput by allocating more radio resources to V2I links when larger observation delays do not have significant impacts on the PC performance. This is further proved by a close examination of the average sum V2I throughput and sum PC performance in Table~\ref{table_Performance_Step2}. In terms of the average sum V2I throughput, MTCC-RRA outperforms AoI-RRA and Delay-RRA algorithms by $20.02\%$ and $20.72\%$, respectively. Meanwhile, MTCC-RRA is slightly inferior to AoI-RRA and Delay-RRA in terms of sum PC performance by $8.14\%$. Therefore, MTCC-RRA can strike a better trade-off between PC performance and V2I throughput compared with the other two RRA algorithms. \par

%It is because AoI-RRA and Delay-RRA algorithms are essentially algorithms that minimize observation delay, and it is evident that the PC with the smallest observation delay all the time without saving communication resources performs best. } 

\subsubsection{Testing results of one episode}
\begin{figure*}[t!]
	\centering 
	\subfigure[MTCC-RRA]{
		\includegraphics[scale=0.255]{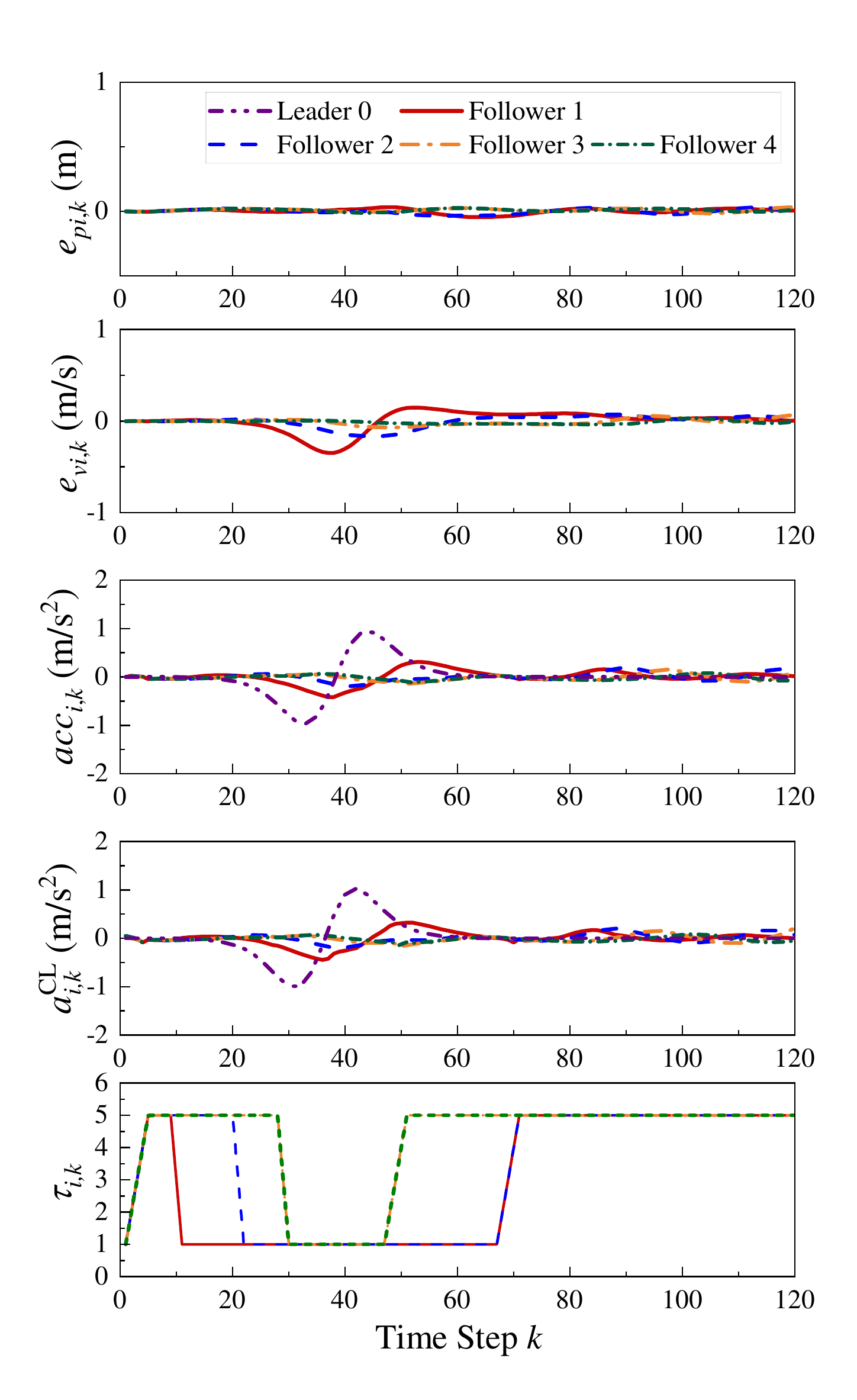}
		\label{MTCC}
	}	
	\subfigure[Delay-RRA]{
		\includegraphics[scale=0.255]{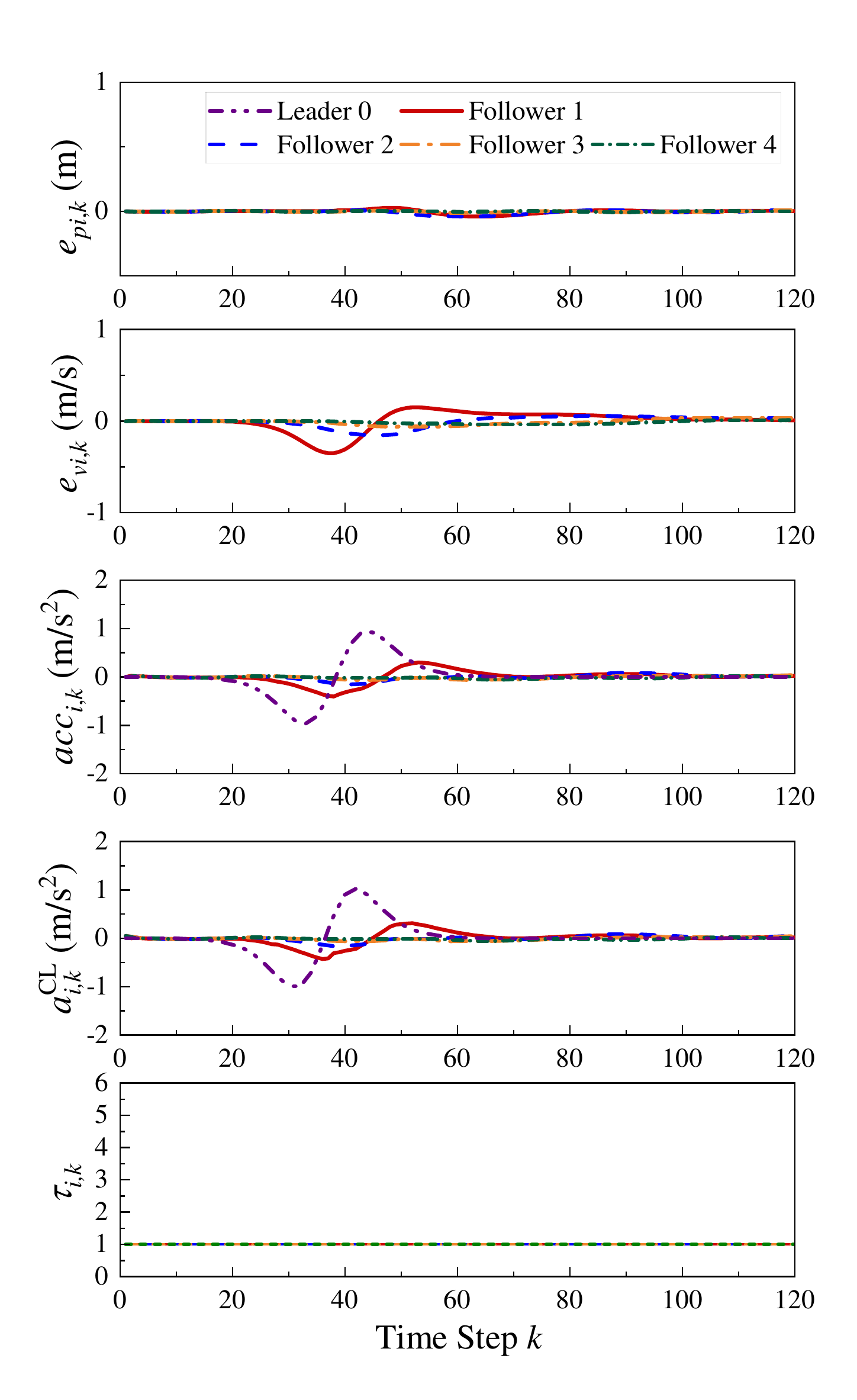}
		\label{Delay_RRA}
	}
	\subfigure[AoI-RRA]{
		\includegraphics[scale=0.255]{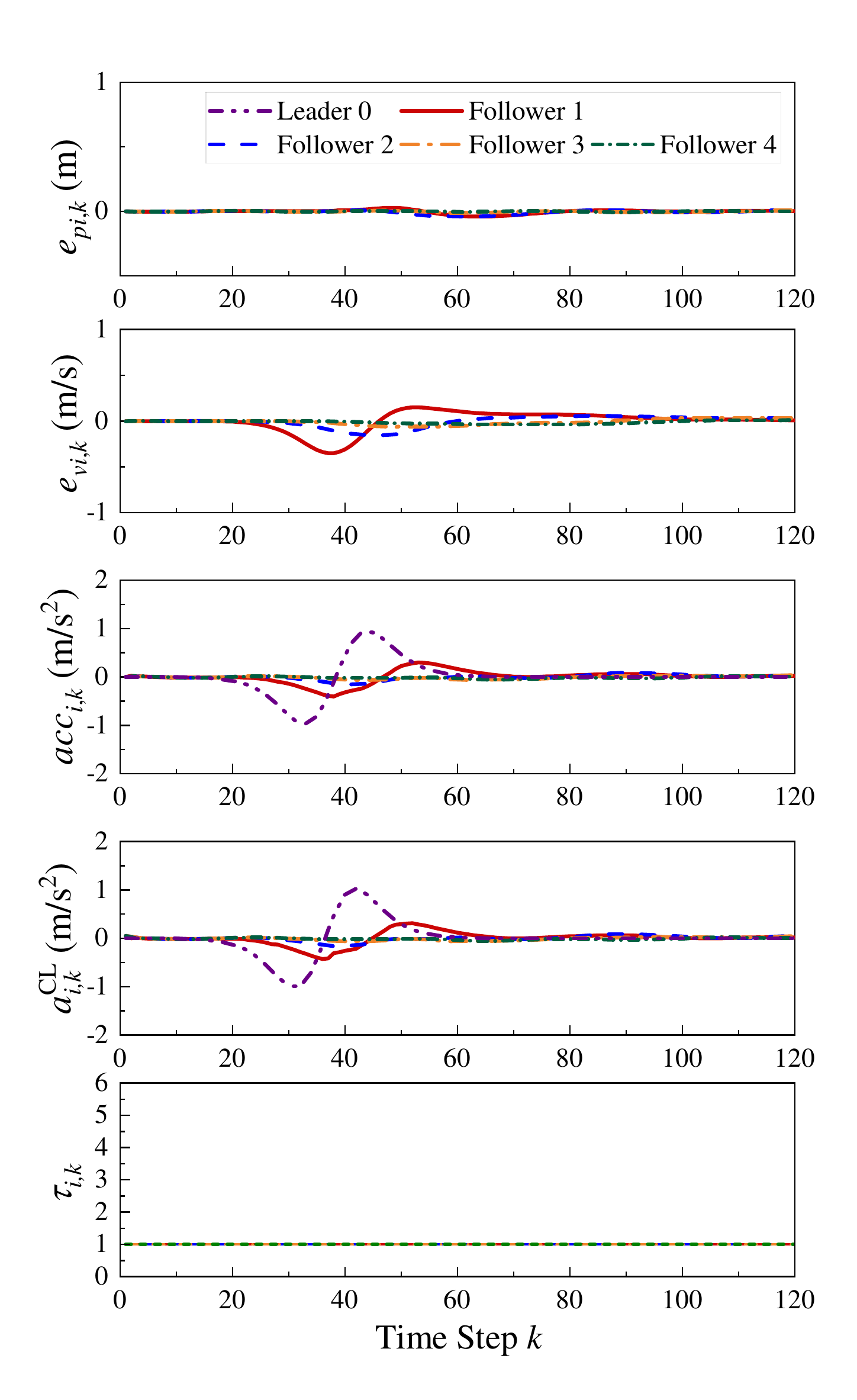}
		\label{AoI_RRA}
	}\par
	\caption{Results of a specific test control episode. The tracking errors $e_{pi,k}$, $e_{vi,k}$, and $acc_{i,k}$ along with the control input $a^{\rm CL}_{i,k}$ and observation delay $\tau_{i,k}$ of each follower $i$ are represented as different curves, respectively.}
	\label{fig_status_Step_2}
\end{figure*}
The key reasons for the superior performance of MTCC-RRA are quantifying the “VoI per control interval” in the reward function and including the control action history in the RRA state. To better understand the positive effects of these design factors, we zoom in to one control episode, where the tracking errors $e_{pi,k}$ and $e_{vi,k}$ of each follower $i\in\mathcal{V}\backslash \{0\}$ as well as the acceleration $acc_{i,k}$ and control input $a^{\rm CL}_{i,k}$ of each vehicle $i\in\mathcal{V}$ for all time steps $k\in\{1,2,\cdots,120\}$ are plotted in Fig.~\ref{fig_status_Step_2} for MTCC-RRA, Delay-RRA, and AoI-RRA. In addition, the observation delay, $\tau_{i,k}$, of each follower $i\in\mathcal{V}\backslash \{0\}$ for each time step $k$ are also plotted. \par

It can be observed that the performance curves of Delay-RRA and AoI-RRA are the same since the minimum observation delay of $1$ control interval is always achieved for both algorithms. Meanwhile, the overall shapes of $e_{pi,k}$, $e_{vi,k}$, $acc_{i,k}$, and $a^{\rm CL}_{i,k}$ curves for MTCC-RRA look very similar to those of Delay-RRA and AoI-RRA, while the $\tau_{i,k}$ curves are quite different. Instead of achieving a constant minimum delay at all time steps, it can be seen that the observation delay of MTCC-RRA for each follower $i\in\mathcal{V}\backslash \{0\}$ is minimum only when the control input $a^{\rm CL}_{i-1,k}$ of its predecessor $i-1$ changes significantly and rapidly. When $a^{\rm CL}_{i-1,k}$ remains relatively constant, the follower $i$ generally has large observation delays, which means that the predecessor $i-1$ refrains from sharing the V2I sub-channels for CAM transmission to support larger V2I throughput. This observation demonstrates that the MTCC-RRA agent has the ability to allocate radio resources based on the VoI per control interval.\par 

As a result, the PC performance of MTCC-RRA is only slightly affected compared to those of Delay-RRA and AoI-RRA. Fig.~\ref{fig_status_Step_2} shows that the convergence speed to steady state for all algorithms are similar, and the PC performance differences of the algorithms are reflected in the oscillations of the tracking errors, acceleration, and control input. After the steady state is reached and the tracking errors are almost zero, the performance curves of MTCC-RRA have slightly larger oscillations compared with those in Delay-RRA and AoI-RRA. As the oscillations are negligible, MTCC-RRA can still ensure driving safety and comfort. Moreover, the amplitudes of the oscillations in $e_{pi,k}$, $e_{vi,k}$, and $acc_{i,k}$ for each follower $i\in\{1,2\}$ are smaller than those of their respective predecessors $i-1$ for all algorithms, which means that the string stability of the platoons can be achieved by MTCC-RRA despite the larger observation delays.  \par

\subsubsection{Convergence properties}
\begin{figure}[tb!]
\centering
\includegraphics[width=0.38\textwidth]{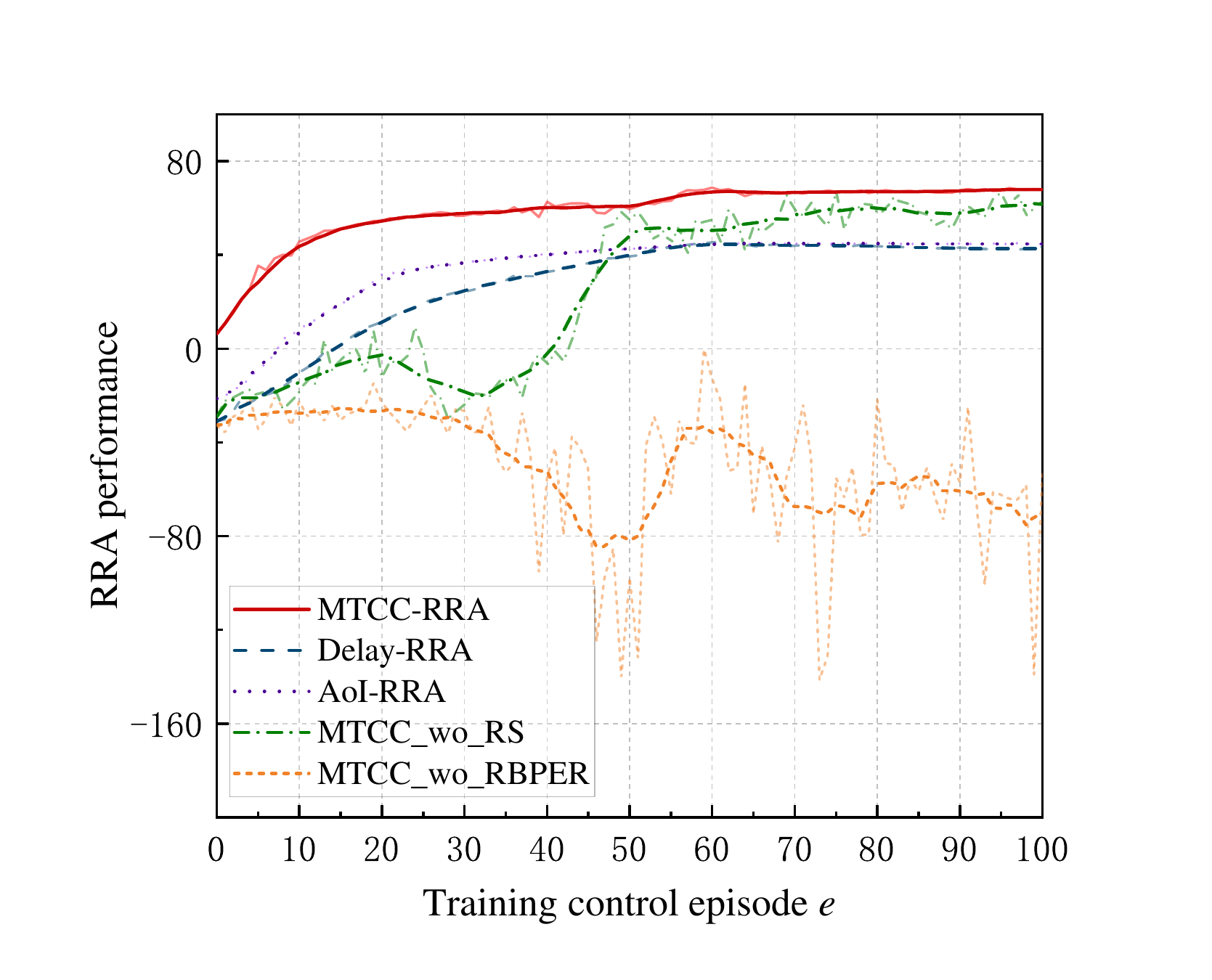}
\caption{RRA Performance during DRL-based RRA algorithms training in one iteration. The vertical axis corresponds to the returns over 1 test episode. The dark curves correspond to smoothed curves and the light color curves correspond to the original curves.}
\label{fig_RRA_convergence}
\end{figure}
The RRA performance of Delay-RRA, AoI-RRA, and MTCC-RRA are evaluated periodically during training by testing in an environment where the trained MTCC-PC policy is used to calculate vehicle trajectories and RRA rewards. Specifically, we run $1$ test control episodes after every $1$ training control episode and calculate the RRA return as the performance for the latest $1$ training episode. The RRA performance as a function of the number of training episodes with MTCC-RRA, Delay-RRA, and AoI-RRA algorithms is plotted in Fig.~\ref{fig_RRA_convergence}. 
It is shown that the convergence rate for all three algorithms are similar, with a ranking of MTCC-RRA $>$AoI-RRA$>$Delay-RRA. \par

\subsection{Performance Comparison of MTCC-RRA, MTCC\_wo\_RS, and MTCC\_wo\_RBPER.}

The RRA performance, average sum V2I throughput, and sum PC performance are reported in Table~\ref{table_Performance_Step2} for MTCC\_wo\_RS and MTCC\_wo\_RBPER, respectively. In terms of RRA performance, MTCC-RRA outperforms MTCC\_wo\_RS and MTCC\_wo\_RBPER algorithms by $16.95\%$ and $132.63\%$, respectively. The superior performance of MTCC-RRA over MTCC\_wo\_RS demonstrates the effectiveness of the reward shaping technique in efficiently distinguishing the multiple agents' credit in the global reward. Moreover, the significant performance gain of MTCC-RRA over MTCC\_wo\_RBPER shows that the RBPER technique is essential in solving the long-range planning with sparse reward problem. The same conclusion can also be drawn from the convergence curves of MTCC\_wo\_RS and MTCC\_wo\_RBPER shown in Fig.~\ref{fig_RRA_convergence}. Both performance curves have large fluctuations, and that of MTCC\_wo\_RBPER is not convergent. \par

\subsection{Performance Comparison of MTCC in one and two iterations}

\begin{figure}[tb!]
\centering
\includegraphics[width=0.45\textwidth]{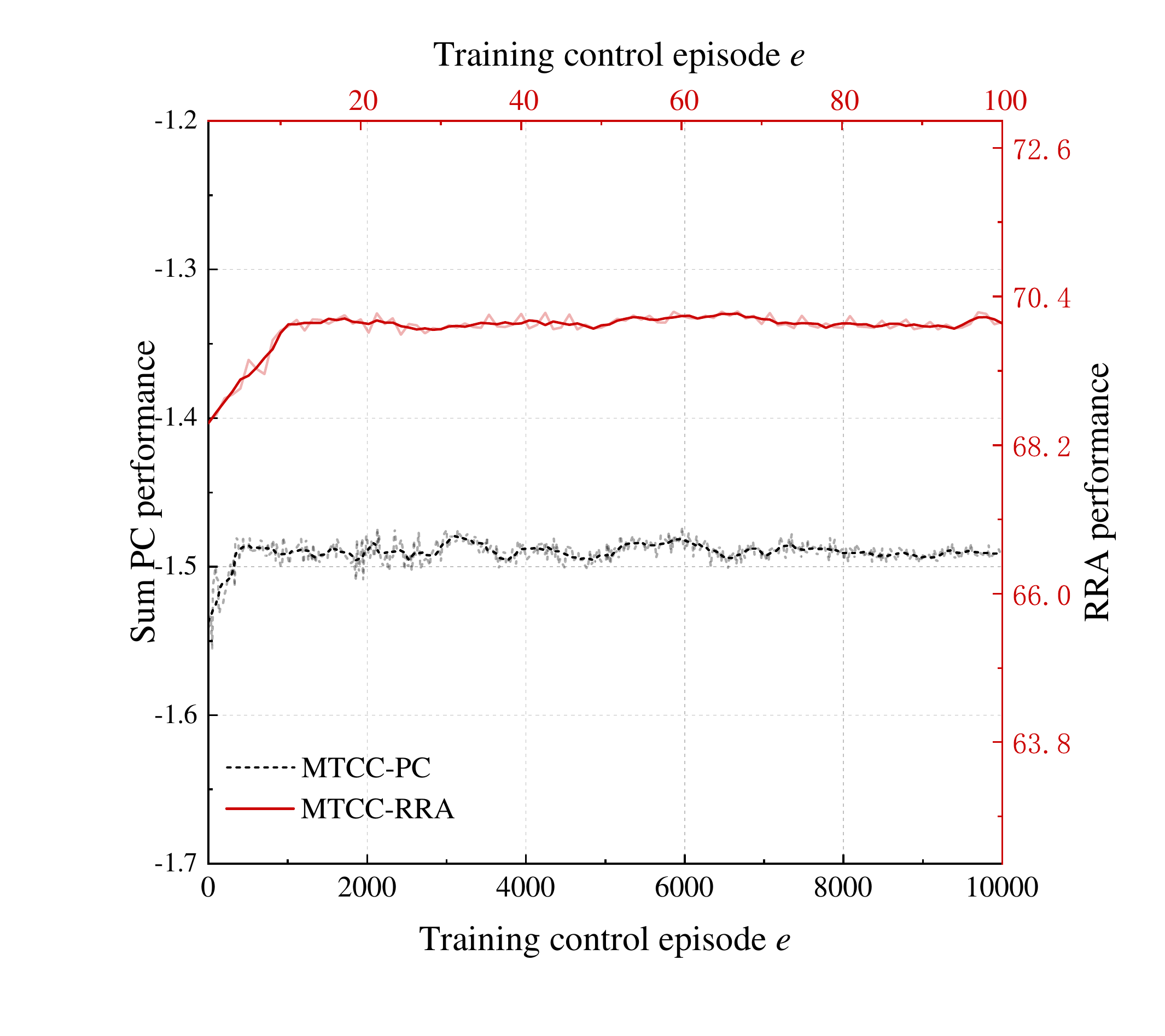}
\caption{Sum PC performance and RRA performance during training in the second iteration.}
\label{fig_convergence_Step3}
\end{figure}

In the above experiments, MTCC is only trained in one iteration, i.e., $Z=1$, in Algorithm \ref{alg1}. In the following, we train MTCC for one additional iteration to observe the resultant performance gain. For the second iteration, the performance curves for the PC policy, i.e., MTCC-PC, during training in Step 1 and for the RRA policy, i.e., MTCC-RRA, during training in Step 2 are given in Fig.~\ref{fig_convergence_Step3}, respectively. The performance curve of MTCC-PC is obtained in a similar way as that in Fig.3 in Part I, except that the MTCC-PC policy is trained and tested under a delayed environment generated by the MTCC-RRA policy learned in Step 2 of iteration 1. Similarly, the performance curve of MTCC-RRA is obtained in a similar way as that in Fig.~\ref{fig_RRA_convergence}, except that the MTCC-RRA policy is trained and tested under the PC environment generated by the MTCC-PC policy learned in Step 1 of iteration 2. From Fig.~\ref{fig_convergence_Step3}, it can be observed that the sum PC performance and RRA performance improve with increasing training episodes in Step 1 and Step 2 of the second iteration, respectively. However, the performance gains are not significant.\par 

Specifically, the sum PC performance of MTCC-PC is around $-1.55$ at the beginning of training and finally converges around $-1.49$ towards the end, with a performance gain of $3.87\%$. The performance gain in Step 1 is due to the fact that the MTCC-PC policy is trained under the observation delay generated by the de facto MTCC-RRA policy instead of the random RRA policy. Similarly, the RRA performance of MTCC-RRA is around $68.4$ at the beginning of training and finally converges around $70.2$ towards the end, with a performance gain of $2.63\%$. The performance gain in Step 2 is because the de facto MTCC-PC policy trained in iteration 2 is adopted instead of that trained in iteration 1 to simulate the PC environment, so that the reward function of MTCC-RRA calculated from the MTCC-PC advantage function can more accurately represent the PC performance degradation due to observation delay.

The above results show that satisfactory policies for MTCC can be learned in one iteration and a second iteration is merely a fine-tuning for small performance improvement. Moreover, it can be anticipated that the performance gain will keep decreasing when training in more iterations.\par

\section{Conclusion}
In this pair of papers, we have studied how to solve the joint optimization of the MTCC problem in the C-V2X system. We have established a unified DRL framework for MTCC and proposed the MTCC-PC algorithm to solve the communication-aware PC sub-problem in Part I of this { two-part} paper. In {this paper}, we have proposed the MTCC-RRA algorithm to solve the control-aware RRA sub-problem. Specifically, the RRA decisions have been made based on the advantage function of the PC model, which provides a fine-grained VoI per control interval. The multi-agent problem and the sparse reward problem in RRA have been tackled with the reward shaping and RBPER techniques, respectively. Finally, we have designed a sample- and computational-efficient training approach to jointly learn the PC and RRA policies iteratively. The experimental results have shown that the proposed MTCC-RRA algorithm outperforms the baseline DRL algorithms, striking a good trade-off between minimizing PC performance degradation due to observation delay and maximizing V2I throughput. Moreover, it is demonstrated that the joint training approach can learn satisfactory PC and RRA policies in only one iteration. In our future work, we will apply the proposed MTCC framework to other autonomous driving tasks such as lane changing and merging.\par

\appendix

\subsection{Proof of Lemma 1}

{As explained in Section II.D, we convert the finite-horizon PC problem to an infinite-horizon one by calculating the TD target $y_{i,K-1}$ using \eqref{TD} for the last control interval $K-1$ instead of setting $y_{i,K-1}$ to be the immediate reward. Therefore, in the following proof, we replace $K$ with $\infty$ on the RHS of \eqref{perf_diff}.\par }

\begin{align}
\label{Proof_Lemma1}
&\mathrm{E}_{\pi^{\rm CM}}\mathrm{E}_{\pi^{\rm CL}_{i}}\left[\sum_{k=0}^{\infty} \gamma^{k} A_{\hat{\pi}^{\rm CL*}_{i}}\left(x_{i,k},\pi^{\rm CL}_{i}(S^{\rm CL}_{i,k})\right)\right] \IEEEnonumber \\ 
& \stackrel{(a)}{=}\mathrm{E}_{\pi^{\rm CM}}\mathrm{E}_{\pi^{\rm CL}_{i}}\bigg[\sum_{k=0}^{\infty} \gamma^{k} \Big(Q_{\hat{\pi}^{\rm CL*}_{i}}\left(x_{i,k},\pi^{\rm CL}_{i}(S^{\rm CL}_{i,k})\right) \IEEEnonumber  \\ 
&-V_{\hat{\pi}^{\rm CL*}_{i}}(x_{i,k})\Big)\bigg] \IEEEnonumber  \\ 
& {\stackrel{(b)}{=} \mathrm{E}_{\pi^{\rm CM}}\mathrm{E}_{\pi^{\rm CL}_{i}}\bigg[\sum_{k=0}^{\infty} \gamma^{k} \Big(\mathrm{E}_{x_{i,k+1}}[R^{\rm CL}_{i,k}+\gamma V_{\hat{\pi}^{\rm CL*}_{i}}(x_{i,k+1})]} \IEEEnonumber  \\ 
&{-V_{\hat{\pi}^{\rm CL*}_{i}}(x_{i,k})\Big)\bigg]} \IEEEnonumber  \\ 
& \stackrel{(c)}{=} \mathrm{E}_{\pi^{\rm CM}}\mathrm{E}_{\pi^{\rm CL}_{i}}\bigg[\sum_{k=0}^{\infty} \gamma^{k} \Big(R^{\rm CL}_{i,k}+\gamma V_{\hat{\pi}^{\rm CL*}_{i}}(x_{i,k+1}) \IEEEnonumber  \\ 
&-V_{\hat{\pi}^{\rm CL*}_{i}}(x_{i,k})\Big)\bigg] \IEEEnonumber  \\ 
&  =  \mathrm{E}_{\pi^{\rm CM}}\mathrm{E}_{\pi^{\rm CL}_{i}}\left[\sum_{k=0}^{\infty} \gamma^{k} R^{\rm CL}_{i,k}-V_{\hat{\pi}^{\rm CL*}_{i}}(x_{i,0})\right] \IEEEnonumber  \\ 
& =\mathrm{E}_{\pi^{\rm CM}}\mathrm{E}_{\pi^{\rm CL}_{i}}\left[\sum_{k=0}^{\infty} \gamma^{k} R^{\rm CL}_{i,k}\right] -\mathrm{E}_{x_{i,0}}\left[V_{\hat{\pi}^{\rm CL*}_{i}}(x_{i,0})\right] \IEEEnonumber \\ 
& \stackrel{(d)}{=} J^{\rm CL}_{i}-\hat{J}^{\rm CL*}_{i} 
\end{align}
\noindent where (a) follows from \eqref{advantage}; (b) follows from the Bellman equation; {(c) follows by merging the expectation $\mathrm{E}_{x_{i,k+1}}$ into the expectation $\mathrm{E}_{\pi^{\rm CM}}\mathrm{E}_{\pi^{\rm CL}_{i}}$;} (d) follows from \eqref{pc_perf}, \eqref{perf_truestate} and the definition of the value function. 
\subsection{Proof of Theorem 1}
\begin{align}
&J^{\mathrm{CM}}=\mathrm{E}_{\pi^{\rm CM}}\mathrm{E}_{\pi^{\rm CL}}\left[\sum_{k=0}^{K-1}\sum_{t=0}^{T-1} \eta^{kT+t} R^{\rm CM}_{(k,t)}\right] \IEEEnonumber \\ 
&\stackrel{(a)}{=}\kappa_1\sum_{m=0}^{M-1}\mathrm{E}_{\pi^{\rm CM}}\left[\sum_{k=0}^{K-1}\sum_{t=0}^{T-1} \eta^{kT+t}r_{m,(k,t)}\right] +\kappa_2 \sum_{i=0}^{N-2}\left(\mathrm{E}_{\pi^{\rm CM}}\right. \IEEEnonumber \\ 
&\left. \mathrm{E}_{\pi^{\rm CL}_{i+1}} \left[\sum_{k=0}^{K-1} \eta^{kT+T-1} A_{\hat{\pi}^{\rm CL*}_{i+1}}\left(x_{i+1,k+1},\pi^{\rm CL}_{i+1}(S^{\rm CL}_{i+1,k+1})\right) \right] \right) \IEEEnonumber \\  
&\stackrel{(b)}{=}\kappa_1\sum_{m=0}^{M-1}\mathrm{E}_{\pi^{\rm CM}}\left[\sum_{k=0}^{K-1}\sum_{t=0}^{T-1} \eta^{kT+t}r_{m,(k,t)}\right]+\kappa_2 \sum_{i=0}^{N-2}\left(\mathrm{E}_{\pi^{\rm CM}}\right. \IEEEnonumber \\ 
&\left.\mathrm{E}_{\pi^{\rm CL}_{i+1}} \left[\sum_{k=0}^{K-1} \gamma^{k} A_{\hat{\pi}^{\rm CL*}_{i+1}}\left(x_{i+1,k+1},\pi^{\rm CL}_{i+1}(S^{\rm CL}_{i+1,k+1})\right) \right] \right) \IEEEnonumber \\ 
&\stackrel{(c)}{=}\kappa_1\sum_{m=0}^{M-1}\mathrm{E}_{\pi^{\rm CM}}\left[\sum_{k=0}^{K-1}\sum_{t=0}^{T-1} \eta^{kT+t}r_{m,(k,t)}\right]+\kappa_2 \sum_{i=0}^{N-2}\left(J^{\rm CL}_{i+1}-\right.\IEEEnonumber \\ 
&\left.\hat{J}^{\rm CL*}_{i+1} \right)\IEEEnonumber \\ 
&\stackrel{(d)}{=}\kappa_1\sum_{m=0}^{M-1}\mathrm{E}_{\pi^{\rm CM}}\left[\sum_{k=0}^{K-1}\sum_{t=0}^{T-1} \eta^{kT+t}r_{m,(k,t)}\right]+\kappa_2 \Delta J^{\mathrm{CL}}.
\end{align}
\noindent where (a) follows from \eqref{comreward_g} and \eqref{comreward_i}; (b) holds if $\eta^{kT+T-1}=\gamma^{k}$; (c) follows from Lemma 1; and (d) follows from \eqref{perf_loss}. Thus it is proved that the expected return as calculated in \eqref{RRA_cumreward} is the same as the optimization objective in \eqref{opti_obj}.

In the above derivation, we assume that $\eta^{kT+T-1}=\gamma^{k}$. Since the PC problem is essentially an infinite horizon problem as discussed in Section II.D, we have 
\begin{align}
\eta=\gamma^{\frac{k}{(k+1)T-1}}  
= \lim\limits_{k\to+\infty} \gamma^{\frac{k}{(k+1)T-1}} = \gamma^{\frac{1}{T}}.
\end{align}

\subsection{Proof of Lemma 2}
If $x_{i,k}$ can be fully determined by $S^{\rm CL}_{i,k}$ with $x_{i,k}=f(S^{\rm CL}_{i,k})$, the relationship between $\hat{\pi}_{i}^{\mathrm{CL}*}(x_{i,k})$ and $\pi_{i}^{\mathrm{CL*}}(S^{\rm CL}_{i,k})$ can be established as 
\begin{equation}
a^{\rm CL*}_{i,k}=\hat{\pi}_{i}^{\mathrm{CL}*}(x_{i,k})=\hat{\pi}_{i}^{\mathrm{CL}*}(f(S^{\rm CL}_{i,k}))=\pi_{i}^{\mathrm{CL*}}(S^{\rm CL}_{i,k})
\end{equation}

\subsection{Proof of Lemma 3}
The system dynamics evolve in discrete time on the basis of forward Euler discretization is
\begin{equation}
	\label{dynamics}
	x_{i,k+1}=g(x_{i,k},a^{\rm CL}_{i,k},a^{\rm CL}_{i-1,k})=Ax_{i,k}+Ba^{\rm CL}_{i,k}+Ca^{\rm CL}_{i-1,k},
	\end{equation}
	\noindent where
	\begin{equation}
	\label{eq13}
	A=\begin{bmatrix}
	1 & T & -h_{i}T & 0 \\
	0 & 1 & -T  & T\\
	0 & 0 & 1-\frac{T}{\tau} &0\\
    0 & 0 &0 &1-\frac{T}{\tau} 
	\end{bmatrix},
	B=\begin{bmatrix}
	0\\
	0 \\
	\frac{T}{\tau} \\
    0
	\end{bmatrix},
	C=\begin{bmatrix}
	0\\
	0 \\
	0 \\
    \frac{T}{\tau}
	\end{bmatrix}.		
	\end{equation}		
Then, 
\begin{align}
\label{iter}
x_{i,k}&=g(x_{i,k-1},a^{\rm CL}_{i,k-1},a^{\rm CL}_{i-1,k-1})\\ \nonumber
&=g(g(x_{i,k-2},a^{\rm CL}_{i,k-2},a^{\rm CL}_{i-1,k-2}),a^{\rm CL}_{i,k-1},a^{\rm CL}_{i-1,k-1})\\ \nonumber
&\dots \\ \nonumber
&=f(x_{i,k-\tau_{i,k}},\{a^{\rm CL}_{i,k'}\}_{k'=k-\tau_{i,k}}^{k-1},\tau_{i,k},\{a^{\rm CL}_{i-1,k'}\}_{k'=k-\tau_{i,k}}^{k-1})\\ \nonumber
&=f(S^{\rm CL}_{i,k},\{a^{\rm CL}_{i,k'}\}_{k'=k-\tau_{i,k}}^{k-1})\\ \nonumber
\end{align}

Therefore, if the sequence of actions of vehicle $i-1$ within the time window of $[k-\tau_{i,k}+1,k]$ is available to vehicle $i$, the current driving status $x_{i,k}$ can be fully determined by the delayed observation $S^{\rm CL}_{i,k}$. \par

{We would like to point out that the state space model in \eqref{dynamics} ignores the random perturbations caused by mechanical noise due to practical system limitations. Therefore, the practical driving status $x_{i,k}$ might slightly deviate from the derived values by \eqref{iter}. However, the sequence of PC actions $\{a^{\rm CL}_{i,k'}\}_{k'=k-\tau_{i,k}}^{k-1}$ of vehicle $i-1$ still play the most important role in predicting the current driving status $x_{i,k}$ based on the delayed observation $S^{\rm CL}_{i,k}$.}

\subsection{Proof of Theorem 2}

From Lemma \ref{lemma2} and Lemma \ref{lemma3}, we can derive that if vehicle $i-1$'s control actions $a^{\rm CL}_{i-1,k'}$ within the period $k'\in[k-\tau_{\mathrm{max}},k-1]$ is available to vehicle $i$ at control interval $k$, the optimal PC policy $\pi_{i}^{\mathrm{CL*}}(S^{\rm CL}_{i,k})$ based on delayed observation $S^{\rm CL}_{i,k}$ performs as well as the optimal policy $\hat{\pi}_{i}^{\mathrm{CL}*}(x_{i,k})$ based on the current driving status $x_{i,k}$ at vehicle $i$. In practice, only the control action $a^{\rm CL}_{i-1,k-\tau_{i,k}}$ of vehicle $i-1$ is available to vehicle $i$ at control interval $k$. Consider the extreme case when $a^{\rm CL}_{i-1,k'}$ is a constant value within period $k'\in[k-\tau_{\mathrm{max}},k-1]$, which means that $a^{\rm CL}_{i-1,k-\tau_{i,k}}$ can accurately represent all of vehicle $i-1$'s control actions within the time window. In this case, $x_{i,k}$ can be fully determined by $S^{\rm CL}_{i,k}$ according to Lemma \ref{lemma3}. Thus, $\hat{\pi}_{i}^{\mathrm{CL}*}(x_{i,k})$ and $\pi_{i}^{\mathrm{CL*}}(S^{\rm CL}_{i,k})$ are equivalent from Lemma \ref{lemma2}. In other words, if the variation of $a^{\rm CL}_{i-1,k'}$ is zero within period $k'\in[k-\tau_{\mathrm{max}},k-1]$, the PC performance at vehicle $i$ is not impacted by the observation delay. Extending to the general case, the smaller the variation of $a^{\rm CL}_{i-1,k'}$ within the period $k'\in[k-\tau_{\mathrm{max}},k-1]$, the more accurate $a^{\rm CL}_{i-1,k-\tau_{i,k}}$ can be used to represent all the control actions of vehicle $i-1$ within the time window, and the smaller the impact of observation delay on PC performance of vehicle $i$.

\bibliographystyle{IEEEtran}
\bibliography{MTCC}
\end{document}